\documentclass[usenatbib,useAMS,usedcolumn]{mnras}
\usepackage{times,epsfig,amsmath}

\hypersetup{pdfauthor={Ewan O'Sullivan},
            pdftitle={Building a cluster: shocks, cavities, and cooling filaments in the group-group merger NGC 6338},
            pdfkeywords={galaxies: elliptical and lenticular, cD;  galaxies: clusters: intracluster medium;  galaxies: groups: individual (WBL 636);  galaxies: individual (NGC6338);  galaxies: active;  X-rays: galaxies: clusters },
            bookmarksnumbered=true}
\pdfoutput=1

\newcommand{\arcm}{\hbox{$^\prime$}}

\newcommand{\degree}{\hbox{$^\circ$}}
\newcommand{\rosat}{\emph{ROSAT}}
\newcommand{\chandra}{\emph{Chandra}}
\newcommand{\xmm}{\emph{XMM-Newton}}
\newcommand{\xmms}{\emph{XMM}}

\newcommand{\arcs}{\mbox{\arcm\arcm}}
\newcommand{\Lx}{\ensuremath{L_{\mathrm{X}}}}
\newcommand{\Tx}{\ensuremath{T_{\mathrm{X}}}}
\newcommand{\Zsol}{\ensuremath{\mathrm{~Z_{\odot}}}}

\newcommand{\Msol}{\ensuremath{\mathrm{~M_{\odot}}}}

\newcommand{\LT}{\ensuremath{\mbox{\Lx :\Tx}}}

\newcommand{\NH}{\ensuremath{N_{\mathrm{H}}}}

\newcommand{\s}{\ensuremath{\mbox{~s}}}
\newcommand{\ps}{\ensuremath{\s^{-1}}}
\newcommand{\cm}{\ensuremath{\mbox{~cm}}}
\newcommand{\pcmsq}{\ensuremath{\cm^{-2}}}
\newcommand{\pcmcu}{\ensuremath{\cm^{-3}}}
\newcommand{\kev}{\ensuremath{\mbox{~keV}}}
\newcommand{\kevcmsq}{\ensuremath{\kev\cm^{2}}}
\newcommand{\km}{\ensuremath{\mbox{~km}}}
\newcommand{\Mpc}{\ensuremath{\mbox{~Mpc}}}
\newcommand{\pMpc}{\ensuremath{\Mpc^{-1}}}
\newcommand{\kmpspMpc}{\ensuremath{\km \ps \pMpc\,}}
\newcommand{\erg}{\ensuremath{\mbox{~erg}}}
\newcommand{\ergps}{\ensuremath{\erg \ps}}
\newcommand{\ergpspcmsq}{\ensuremath{\erg \ps \pcmsq}}
\newcommand{\kmps}{\ensuremath{\km \ps}}

\newcommand{\Ho}{\ensuremath{H_\mathrm{0}}}

\newcommand{\Dtf}{\ensuremath{D_{\mathrm{25}}}}

\newcommand{\gtsim}{\,\rlap{\raise 0.5ex\hbox{$>$}}{\lower 1.0ex\hbox{$\sim$}}\,}

\begin{document}

\title[
The group-group merger NGC~6338
]{
Building a cluster: shocks, cavities, and cooling filaments in the group-group merger NGC 6338
}

\author[Ewan O'Sullivan et al.]
{Ewan O'Sullivan$^{1}$\thanks{eosullivan@cfa.harvard.edu},
  Gerrit Schellenberger$^{1}$, D.~J. Burke$^{1}$, Ming Sun$^{2}$, Jan M. Vrtilek$^{1}$, \newauthor Laurence P. David$^{1}$ and Craig Sarazin$^{3}$\\
$^1$ Harvard-Smithsonian Center for Astrophysics, 60 Garden
  Street, Cambridge, MA 02138, USA \\
$^2$ Department of Physics \& Astronomy, University of Alabama in Huntsville, Huntsville, AL 35899, USA\\
$^3$ Department of Astronomy, University of Virginia, 530 McCormick Road, Charlottesville, VA 22904-4325, USA\\
}

\date{Accepted 2019 June 17; Received 2019 June 17; in original form 2019 May 1}

\pagerange{\pageref{firstpage}--\pageref{lastpage}} \pubyear{2011}

\label{firstpage}

\maketitle

\begin{abstract}
We present deep \chandra, \xmm, Giant Metrewave Radio Telescope and H$\alpha$ observations of the group-group merger NGC~6338. X-ray imaging and spectral mapping show that as well as trailing tails of cool, enriched gas, the two cool cores are embedded in an extensive region of shock heated gas with temperatures rising to $\sim$5~keV. The velocity distribution of the member galaxies show that the merger is occurring primarily along the line of sight, and we estimate that the collision has produced shocks of Mach number $\mathcal{M}$=2.3 or greater, making this one of the most violent mergers yet observed between galaxy groups. Both cool cores host potential AGN cavities and H$\alpha$ nebulae, indicating rapid radiative cooling. In the southern cool core around NGC~6338, we find that the X-ray filaments associated with the H$\alpha$ nebula have low entropies ($<$10\kevcmsq) and short cooling times ($\sim$200-300~Myr). In the northern core we identify an H$\alpha$ cloud associated with a bar of dense, cool X-ray gas offset from the dominant galaxy. We find no evidence of current jet activity in either core. We estimate the total mass of the system and find that the product of this group-group merger will likely be a galaxy cluster.
\end{abstract}


\section{Introduction}
\label{sec:intro}

At the core of hierarchical models of structure formation is the prediction that massive gravitationally bound systems form through the merger of many less massive progenitors. Evidence of such mergers is seen at all scales from individual galaxies \citep[e.g.,][]{Bundyetal09} to the most massive galaxy clusters \citep[e.g.,][]{Schellenbergeretal19}. Mergers have a significant impact on the hot gaseous components of galaxy groups and clusters, injecting energy and helping to mix gas of different temperatures and metallicities \citep[e.g.,][]{MarkevitchVikhlinin07}. The \chandra\ and \xmm\ X-ray observatories have allowed the study of numerous cluster mergers in great detail, revealing encounters ranging from high-velocity collisions driving powerful shocks \citep[e.g.,][]{Markevitch02,Markevitchetal05,Russelletal12b,Dasadiaetal16} to tangential mergers whose signature is the excitation of ``sloshing'' oscillations in the intra-cluster medium \citep[ICM, e.g.,][]{Roedigeretal11,Johnsonetal12,Paterno-Mahleretal13}. X-ray studies of merging galaxy groups are less common, owing to the lower luminosity of these systems, and most focus on low-energy interactions \citep[e.g.,][]{Machaceketal05b,Kraftetal06,Machaceketal10,Machaceketal11,Roedigeretal12,Gastaldelloetal13,OSullivanetal14a}. Only a handful of studies have shown evidence of shock heating by group-group mergers \citep[e.g.,][]{Randalletal09,Russelletal14}. Galaxy groups are far more common than more massive galaxy clusters \citep{Tully87} and are the environment in which most galaxies reside \citep{Ekeetal06}, and it is therefore desirable to understand the range of impacts which mergers may have on their properties. 

Groups are also a key environment for the study of AGN feedback. Observations over the past two decades have established that the ICM of cool core clusters and groups is thermally regulated by the jets of cluster-central FR-I radio galaxies \citep[e.g,][]{Fabian12,McNamaraNulsen12}. X-ray observations reveal cavities in the ICM coincident with the radio lobes of these galaxies, and the enthalpy associated with them has been shown to be sufficient to balance radiative cooling \citep[e.g.,][]{Birzanetal08,Cavagnoloetal10,OSullivanetal11b,Birzanetal12}. AGN outbursts can also drive shocks which may have a significant heating impact in groups \citep[e.g.,][]{Randalletal15}. 

The supermassive black holes (SMBHs) of these radio galaxies are thought to be fed by material cooling from the ICM. Evidence of this cooled material is observed, in the form of molecular  \citep[e.g.,][]{Salomeetal06,Salomeetal11,Davidetal14,Russelletal16,Vantyghemetal16,Russelletal17a,Russelletal17b,Vantyghemetal17,Temietal17} and H$\alpha$-emitting (10$^4$~K) ionized gas \citep[e.g.,][]{Fabianetal03b,Crawfordetal05,Crawfordetal05b,McDonaldetal10,McDonaldetal11,Lakhchauraetal18}, often found in the form of filamentary structures correlated with AGN cavities. Studies have shown that jet power is correlated with the mass of molecular gas \citep{Babyketal18} and (tentatively) the H$\alpha$ luminosity \citep{Lakhchauraetal18} in early-type galaxies, showing that AGN feedback is closely linked to the presence of cooled material. The filamentary structures are thought to form from ICM gas that becomes thermally unstable \citep{Sharmaetal12,McCourtetal12,Gasparietal13,LiBryan14a,LiBryan14b}, either in situ or as the result of disturbances caused by the AGN or gas motions \citep[e.g.,][]{McNamaraetal16,Prasadetal17,Voitetal17,Hoganetal17,Gasparietal18}. As with mergers, much of the study of these phenomena has focused on the more luminous galaxy clusters, and there are as yet relatively few examples of detailed studies of H$\alpha$ filaments and molecular gas in galaxy groups.

\begin{figure*}
\includegraphics[width=\columnwidth,trim=50 140 50 140,clip]{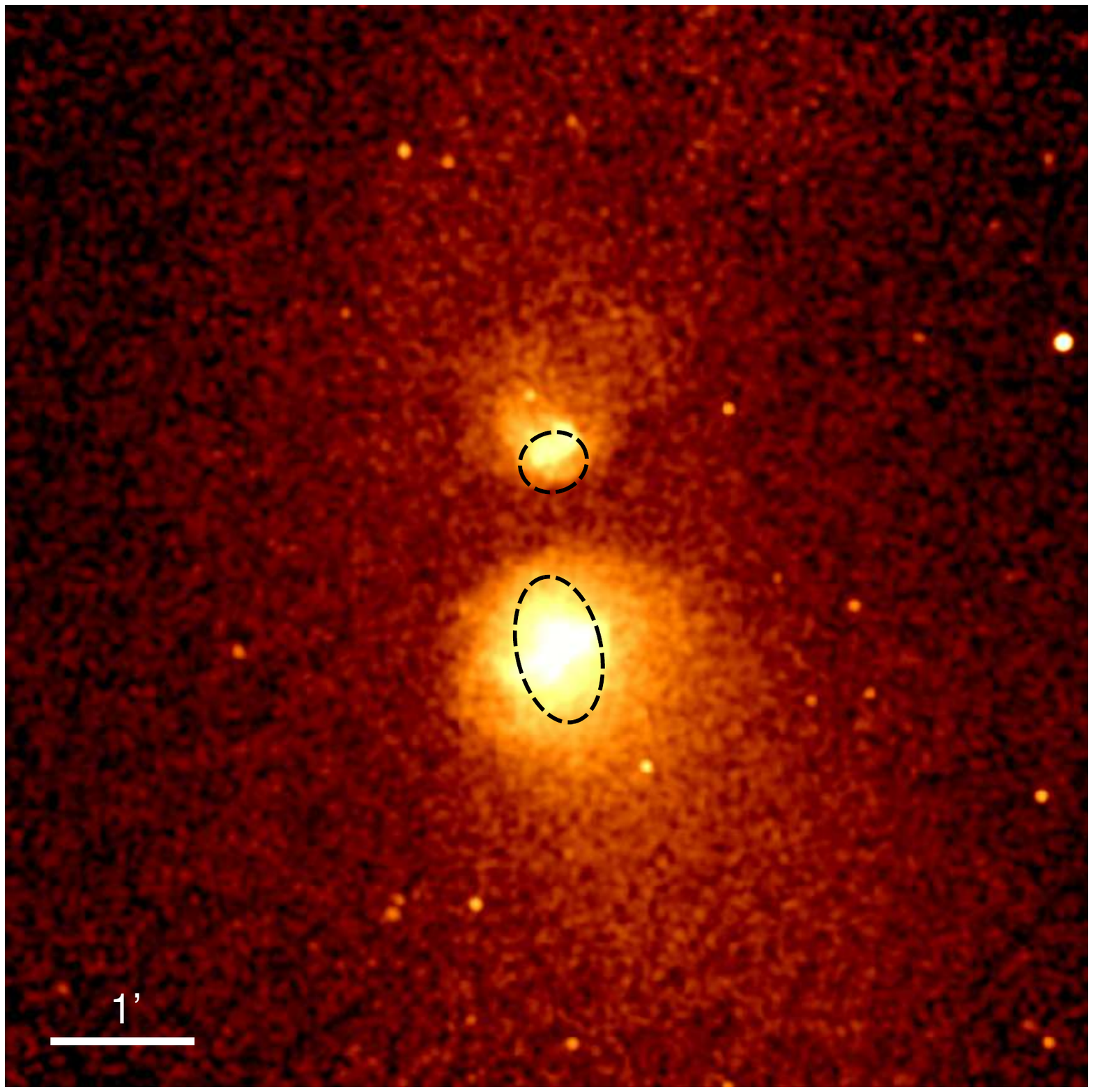}
\includegraphics[width=\columnwidth,trim=50 140 50 140,clip]{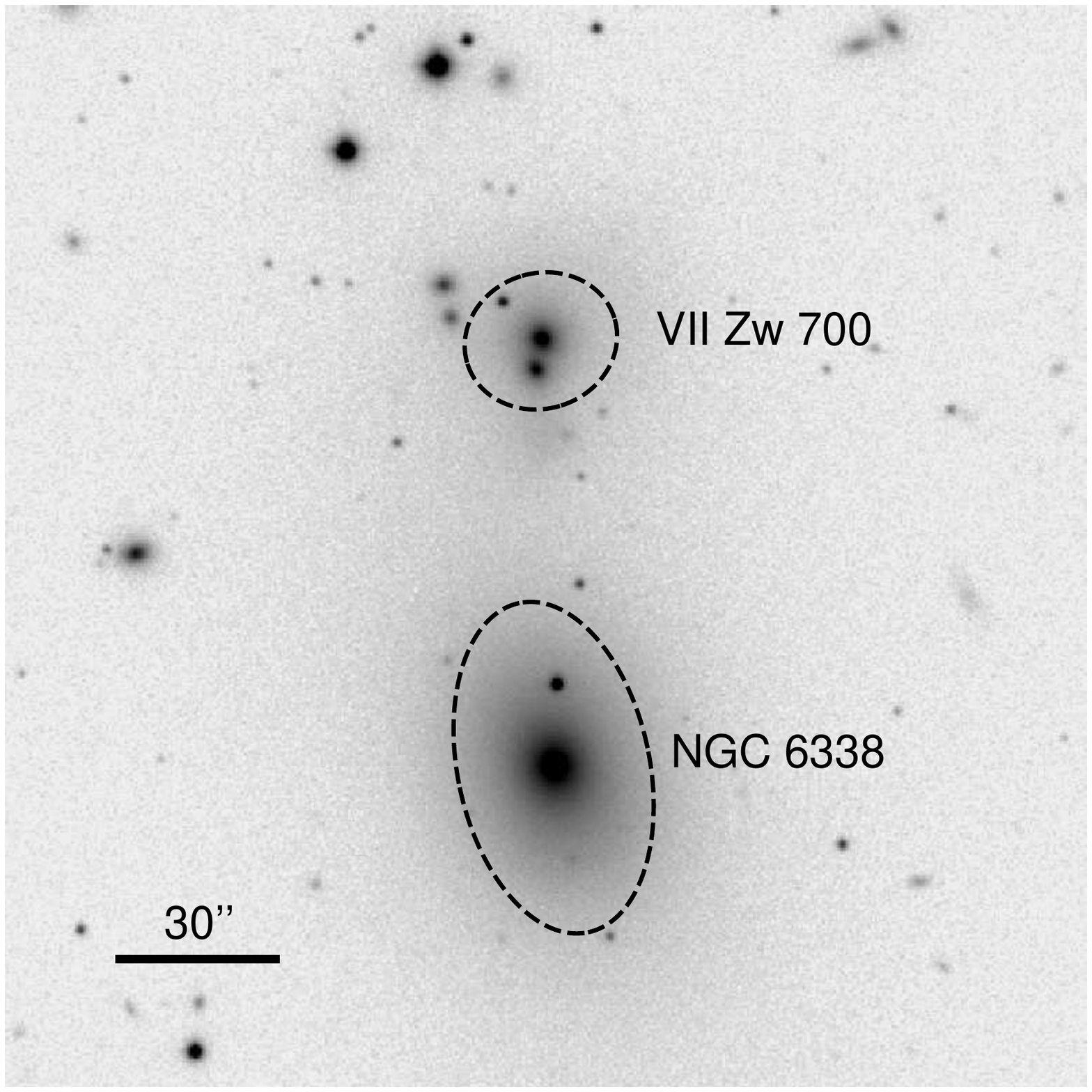}
\caption{\label{fig:introims}\textit{Left}: Merged, exposure corrected \chandra\ 0.5-2~keV image of the NGC~6338 group, smoothed with a 3\arcs\ Gaussian. \textit{Right}: SDSS $i$-band image of the system. The \Dtf\ ellipses of the dominant galaxies, NGC~6338 and VII~Zw~700 are marked by black dashed lines in both panels.}
\end{figure*}

The NGC~6338 group shows features suggestive of both rapid cooling and merging. X-ray observations reveal extended diffuse emission with two peaks. The brighter southern peak is centred on the most optically luminous galaxy NGC~6338, while the northern peak is centred close to a galaxy pair, VII~Zw~700, which is dominated by the elliptical MCG+10-24-117. NGC~6338 and MCG+10-24-117 are separated by $\sim$80\arcs\ ($\sim$42~kpc) in projection, and $\sim$1400\kmps\ in velocity. The large velocity difference and double-peaked X-ray distribution suggests a merging system. Figure~\ref{fig:introims} shows optical and X-ray images of the group. \citet{DupkeMartins13} noted the presence of a density discontinuity consistent with a cold front on the south side of the northern peak, and suggested that the northern peak is the core of an infalling system. More recently, \citet{Wangetal19} used \chandra\ ACIS-I and galaxy velocity data to classify the system as a head-on merger with a large component of its velocity along the line of sight. They identified cold fronts associated with both surface brightness peaks, and noted the presence of high temperatures, likely caused by merger shocks, between and around the two peaks.

NGC~6338 was classed as an S0 by the Uppsala and RC3 galaxy catalogues \citep{RC3,Nilson73}, but examination of Sloan Digital Sky Survey data show it to be a prolate elliptical \citep{NairAbraham10}, a classification confirmed by the finding that the stellar component rotates around the major axis \citep{Gomesetal16b}. \textit{Hubble} Space Telescope narrow band imaging shows H$\alpha$ emission from two compact clouds in the central arcsecond of the galaxy, and three filaments extending out to $\sim$13\farcs5 along the minor axis of the galaxy \citep{Marteletal04}. Integral field spectroscopy shows that these filaments are not rotating with the stellar component \citep{Gomesetal16b}, and provides a total H$\alpha$ flux of 1.78$\times$10$^{-14}$\ergpspcmsq\ (L$_{H\alpha}$=2.82$\times$10$^{40}$\ergps\ for our adopted distance). A short \chandra\ X-ray observation showed that the H$\alpha$ filaments have X-ray counterparts \citep{Pandgeetal12} and revealed a pair of small X-ray cavities \citep{Dongetal10} aligned perpendicular to the filaments. Estimates of the enthalpy of these cavities suggest that they are too small to balance cooling in the central 10~kpc over the past $\sim$10~Myr \citep{Pandgeetal12}. However, these structures suggest that cooling from the intra-group medium (IGM) may be the source of the H$\alpha$ emission, and may have fuelled AGN activity in the recent past.

NGC~6338 hosts a flat-spectrum radio point source coincident with the optical nucleus, and is classed as an FR0 radio galaxy \citep{Torresietal18} and a LINER \citep{Gomesetal16b}. The galaxy has been included in catalogues of blazars and BL Lacs \citep[e.g.,][]{Massaroetal09,Mingalievetal14}, but optical studies show this to be a mistake \citep[e.g.,][]{Caccianigaetal02}. The classification seems to have been based on the flat radio spectrum and an early X-ray detection which did not distinguish between diffuse and nuclear emission (\citealt{Marchaetal01}; see also \citealt{Greenetal17}). \citet{Pandgeetal12} report a nuclear X-ray point source, but \citet{Torresietal18} are only able to place an upper limit on a nuclear powerlaw component, F$_{2-10keV}$$<$1.9$\times$10$^{-14}$\ergpspcmsq.

The \textit{Planck} observatory detected the Sunyaev-Zel'dovitch (SZ) signal of the system with a signal-to-noise ratio (S/N) of 5.5 \citep{Planckcollab16}. The total mass of the system was estimated to be [1.01$\pm$0.12]$\times$10$^{14}$\Msol, placing it on the group/cluster boundary.

In this paper, we use a combination of new, deep \chandra\ and \xmm\ X-ray
observations, H$\alpha$ imaging and spectra, and archival GMRT radio data
to investigate the dynamical state of the system and the evidence of
cooling and AGN feedback in the dominant galaxies. The paper is organised
as follows. Section~\ref{sec:obs} describes the observations and our data
reduction, Section~\ref{sec:res} describes the results drawn from the
observations, in Section~\ref{sec:disc} we discuss the dynamical state of
the group and the evidence of cooling and feedback in the two cores, and in
Section~\ref{sec:conc} we summarise our results and list our conclusions.
Throughout this paper we adopt \Ho=70\kmpspMpc, $\Omega_M=0.3$, and
$\Omega_{\Lambda}=0.7$. The redshift of the cluster is taken to be
0.027427, and the resulting angular size and luminosity distances are
$D_A$=109~Mpc and $D_L$=115~Mpc respectively. This gives an angular scale
of 0.528~kpc/\arcs. We report 1$\sigma$ uncertainties unless otherwise
stated in the text.

\section{Observations and Data Reduction}
\label{sec:obs}

NGC~6338 has been observed several times by the \chandra\ and \xmm\ X-ray observatories (see Table~\ref{tab:obs}). Details of instrument-specific reduction and analysis are given in the following sections. X-ray spectral fitting was performed using \textsc{Xspec} 12.9.1u \citep{Arnaud96}. We adopt a Galactic hydrogen column of \NH=2.23$\times$10$^{20}$\pcmsq, drawn from the Leiden-Argentine-Bonn survey \citep{Kalberlaetal05} throughout. The estimated column including molecular hydrogen \citep{Willingaleetal13} is $<$10\% greater, and testing shows that using this higher value has no significant impact on our fits. We adopted the solar abundance ratios of \citet{GrevesseSauval98}. Surface brightness modelling was performed in \textsc{ciao sherpa} v4.10 \citep{Freemanetal01}.

\subsection{Chandra}

NGC~6338 was initially observed by \chandra\ ACIS-I early in the mission, followed by a series of ACIS-S observations during 2017. Table~\ref{tab:obs} gives details of the observation dates, setup and cleaned exposures. A summary of the \chandra\ mission is provided in \citet{Weisskopfetal02}. We processed observations using \textsc{ciao} 4.10 and CALDB 4.7.9. This includes the most recent corrections to the ACIS contamination model, including corrections for the reduced rate of contamination buildup during 2017 \citep{Plucinskyetal18}. Our reduction followed the approach laid out in the \chandra\ analysis threads\footnote{http://cxc.harvard.edu/ciao/threads/index.html} and \citet{OSullivanetal17}.

\begin{table}
\caption{\label{tab:obs}Summary of the X-ray observations. For \xmms\ observing mode indicates that the EPIC-PN operated in either Full Frame (FF) or Extended Full Frame (EFF) mode, and cleaned exposures for the EPIC-MOS/PN are given in the last column.}
\begin{center}
\begin{tabular}{lcccc}
\hline
ObsID & Observation & Instrument & Observing & Cleaned \\
      & date        &            & mode      & exposure (ks) \\
\hline
\multicolumn{5}{l}{\textit{Chandra}}\\
4194  & 2003 Sep 17 & ACIS-I & VFAINT & 46.6 \\
18892 & 2017 Jun 24 & ACIS-S & VFAINT & 11.9 \\
18893 & 2017 Jul 21 & ACIS-S & VFAINT & 44.6 \\
19934 & 2017 Jul 12 & ACIS-S & VFAINT & 28.7 \\
19935 & 2017 Jun 05 & ACIS-S & VFAINT & 35.1 \\
19937 & 2017 Jun 08 & ACIS-S & VFAINT & 19.8 \\
20089 & 2017 Jun 11 & ACIS-S & VFAINT & 18.4 \\
20104 & 2017 Jun 21 & ACIS-S & VFAINT & 14.9 \\
20112 & 2017 Jul 13 & ACIS-S & VFAINT & 41.5 \\
20113 & 2017 Jul 15 & ACIS-S & VFAINT & 13.8 \\
20117 & 2017 Jul 23 & ACIS-S & VFAINT & 12.9 \\
\multicolumn{4}{r}{Total \textit{Chandra} exposure:} & 288.2 \\
\multicolumn{5}{l}{\textit{XMM-Newton}}\\
0741580101 & 2014 Dec 04 & EPIC & FF  & 11.8/9.3 \\
0792790101 & 2016 Oct 12 & EPIC & EFF & 55.9/44.0 \\
\hline
\end{tabular}
\end{center}
\end{table}

Periods of high (flaring) background were filtered using the \textsc{lc\_clean} script. As the ACIS-S observations were individually short and performed over a one month period, we treated them as a single observation for flare filtering. Very faint mode filtering was applied to all observation and background event files. We used the standard set of \chandra\ blank-sky background files to create background spectra and images where needed, normalizing to the observations using the 9.5-12~keV count rate. 

All observations were reprojected onto a common tangent point, and combined images and exposure maps created using \textsc{reproject\_obs} and \textsc{merge\_obs}. Using point sources detected in all observations, we confirmed that there were no significant astrometry errors affecting the final images. For image analysis we used the combined images, typically in the 0.5-2~keV band. Spectral extraction was performed on each observation separately, and the ACIS-S observations combined into single sets of spectra and responses for each region. 

Point sources were identified using \textsc{wavdetect} on the combined 0.5-7~keV image and associated exposure map. A combined map of the point-spread function (PSF) size was created from individual maps of the 2.3~keV 90\% encircled energy fraction in each observation, weighted using the exposure maps. False detections associated with gas structures in the group cores were identified by eye and removed; no genuine point sources were identified in either core. Point sources were generally excluded from all further analysis (excluding at least 90\% of the flux from each source) and where necessary the resulting space was refilled using the \textsc{dmfilth} task.

\subsection{XMM-Newton}
\xmms\ has observed NGC~6338 twice, a short ($\sim$10~ks) observation in 2014, and a longer ($\sim$70~ks) exposure in 2016. The former was not significantly affected by background flaring, but high background makes roughly one third of the latter unusable. Table~\ref{tab:obs} provides details of both observations.

We reduced and analysed data from the European Photon Imaging Cameras (EPIC) for both observations following the approach described in \citet{OSullivanetal17}. Analysis was performed using the \xmms\ Science Analysis System (\textsc{sas} 17.0.0). An initial analysis was performed using a three-band background filter and scaled blank-sky background files, including imaging and the creation of spectral maps. The longer observation was then processed with the \xmms\ Extended Source Analysis Software (\textsc{esas}), spectra were extracted along a number of radial profiles, and fitting was performed using a background model as described in \citet{Snowdenetal04}. For ObsID 0792790101 the EPIC-MOS1 CCDs~3 and 6 were inactive owing to micrometeorite damage, and we excluded rows 0-149 of CCD~4 given their increased noise levels. The \textsc{mos-filter} task found CCD 5 of EPIC-MOS2 to be in an anomalous state, and it was also excluded from further analysis. Point sources were identified using the \textsc{cheese} task.

\subsection{Spectral mapping}
We take a common approach to the creation of spectral maps from the \xmms\ and \chandra\ data. A regular grid of fixed size map pixels is established covering the area of interest. The map pixel scales chosen were 15\arcs\ for \xmms\ and 5\arcs\ or 2\arcs\ for \chandra. Circular spectral extraction regions are then defined, centred on each map pixel, with radii chosen to include a fixed number of net counts, or to achieve a desired signal-to-noise ratio (S/N). We created maps with 2000 net counts for \xmms\ and 1500 for \chandra\ (equivalent to S/N=25-40), as well as a S/N=50 \chandra\ map to examine the abundance distribution. The radius of the regions varies with surface brightness, with smaller regions in the bright cores, and larger regions in the outskirts, up to a set maximum radius. Spectral extraction regions are defined to be no smaller than the map pixels, but can be larger than them, in which case they can overlap and are not necessarily independent. The resulting maps are analogous to adaptively smoothed images, with low surface brightness regions being more heavily smoothed. Table~\ref{tab:maps} lists the pixel scale and other parameters for our maps.

\begin{table*}
\caption{\label{tab:maps}Parameters and uncertainties for spectral maps}
\begin{center}
\begin{tabular}{lccccccc}
\hline
Satellite & pixel scale & criterion & extraction radii & \multicolumn{4}{c}{1$\sigma$ statistical uncertainties} \\
          &             &           &                   & \multicolumn{2}{c}{kT} & \multicolumn{2}{c}{Abundance} \\
          & (arcsec)    &           & (arcsec)          & cool & hot & cool & hot \\
\hline
\xmm\     & 15     & 2000 net ct. & 8-86 & 3-8\% & 17\% & - & - \\
\chandra\ & 5 or 2 & 1500 net ct. & 5-37 & 2-5\% & 15\% & - & - \\
          & 5      & S/N=50       & 4-59 & 1-5\% & 7\%  & 10-20\% & 50\% \\
          & contbin& S/N=30       & -    & 2-5\% & 15\% & - & - \\
\hline
\end{tabular}
\end{center}
\end{table*}

Source and blank-sky spectra were extracted from every \chandra\ or \xmms\ observation covered by each region. For \chandra\ spectra, responses were created for every spectrum, with the resolution of the weighting maps used in calculating the response matrix function (RMF) reduced by a factor of 4 to save time. For \xmms\ spectra a 13$\times$13 grid of responses was created and spectra were assigned a set of responses based on which grid area their center fell within. ACIS-S spectra and responses for each region were combined into single spectra. 

Spectra were then simultaneously fitted (ACIS-I with ACIS-S, or MOS with PN) with single temperature absorbed APEC thermal plasma models \citep{Smithetal01}. The best fitting temperature and abundance from each fit were then used to populate the corresponding map pixel, producing 2-dimensional maps of these properties. Typical 1$\sigma$ uncertainties on fitted parameters are given in table~\ref{tab:maps}. Note that uncertainties are generally smaller in the cool cores than in the hottest parts of the system. We follow \citet{Rossettietal07} in defining a pseudo-density as the square root of the best-fitting normalisation per unit area, and combine this with the temperature map to create maps of pseudo-entropy and -pressure. While these are not measures of the true gas properties, they provide useful hints to the state of the gas which can then be followed up with standard spectral analysis.

In addition, we also created \chandra\ maps using the contour binning approach of \citet{Sandersetal06}. This adaptively bins the area of interest, linking pixels with similar surface brightness to form regions with a desired signal-to-noise ratio, in our case S/N=30. Spectra and (full resolution) responses are then extracted for each region, and fitting is performed as for the other maps. Typical errors on temperature were similar to the ``fixed grid'' \chandra\ maps, $\sim$2-5\% in the cool cores rising to $\sim$15\% in the hottest regions. Contour binning uses independent regions whose boundaries tend to follow surface brightness structures. It is therefore well suited to tracing fronts and edges in the surface brightness distribution, particularly when the temperature and abundance distribution is correlated with these features. By contrast, our adaptive-smoothing-like approach may have an advantage in regions where this assumed correlation does not hold.

Comparison of the \xmms\ and \chandra\ maps shows generally good agreement in areas of overlap. Comparison of the ``fixed grid'' \chandra\ maps with the contour binning maps and with radial spectral analysis also shows good agreement. We are therefore confident that the maps give a generally reliable indication of the 2-dimensional distribution of gas properties in the system.

\subsection{H$\alpha$ data}

NGC~6338 was observed on Sep. 16, 2009 (UT) with the Seaver Prototype Imaging
camera (SPIcam) on the Apache Point Observatory (APO) 3.5m telescope. The night was photometric and
the seeing was $\sim$1.2\arcs.  Two narrow-band filters were used, NMSU
673.6/8 ($\lambda_0$ = 6736~\AA, FWHM = 80~\AA) for the H$\alpha$+[NII]
lines and NMSU 665/8 ($\lambda_0$ = 6650~\AA, FWHM = 80~\AA) for the
continuum.  There is one 673.6/8 observation with an exposure time of 10
minutes.  There are two 665/8 observations, each with 6 minutes.  Each
image was reduced using the standard procedures with the IRAF package.  The
pixels were binned 2$\times2$, for a scale of 0.28\arcs\ per pixel.  The
spectroscopic standard was BD+28~4211. The H$\alpha$+[NII] net image was
derived, with the continuum subtracted by nulling the galaxy outskirts beyond
the central emission line nebula.

We also took long-slit spectra of NGC~6338 and VII~Zw~700 with the Dual Imaging
Spectrograph (DIS) on the APO 3.5m telescope on Sep. 17, 2017. The night
was partially photometric, seeing was $\sim$1.2\arcs\ and the slit width was
2.0\arcs. The B1200/R1200 grating was used, giving a spectral
resolution of 0.62~\AA/pix in the blue channel and 0.58~\AA/pix in the red
channel. Two slit positions were observed, one across NGC~6338 (slit angle 50.67\degree\ north from due east) for three
10-min exposures and another across VII~Zw~700 (slit angle 42.8\degree) for three 10-min exposures.

\subsection{GMRT radio}
\label{sec:radio}

The Giant Metrewave Radio telescope (GMRT) has observed NGC~6338 several times at different frequencies. We chose to analyse observations at 147, 333, and 1388~MHz, providing broad spectral coverage and including the deepest observations. The data were reduced using the \textsc{spam} pipeline \citep{Intemaetal09}. This flags bad data, applies flux and phase calibrations, and corrects for time and sky-position dependent variation in the phase solution. Table~\ref{tab:radio} provides the project codes for each dataset, observation parameters, and the details of the beam and noise level in the full resolution images.

\begin{table*}
\caption{\label{tab:radio} Summary of the GMRT radio observations.}
\begin{center}
\begin{tabular}{lcccccc}
\hline
Project code & Observation date & Frequency & Bandwidth & On-source time & HPBW, PA & rms noise \\
 & & (MHz) & (MHz) & (min) & (\arcs$\times$\arcs, \degree) & ($\mu$Jy/beam) \\
\hline
21\_060 & 2011 Nov 26 & 1388 & 33 & 312 & 2.69$\times$2.29, -56.41 & 30 \\
06EFA01 & 2004 Sep 10 & 333  & 16 & 18 & 25.84$\times$9.24, 78.85 & 410 \\
TGSS R53D67 & 2011 May 26 & 147 & 17 & 15 & 25$\times$25, 0.0 & 3230 \\
\hline
\end{tabular}
\end{center}
\end{table*}

We used the calibrated $uv$ products from the SPAM pipeline to perform the imaging in \textsc{casa} \citep{McMullinetal07} via the \textsc{tclean} task. We applied a Briggs visibility weighting scheme with the robust parameter set to 1, and a minimum cut for $uv$ distances of 1~k$\lambda$. For the deconvolution we chose the multi-scale algorithm \citep{cornwell08} to model extended emission, as well as the wproject gridder \citep{cornwelletal08} with 259 planes.

The previously known radio source in NGC~6338 is visible in the full resolution images at all three frequencies. At both 147~MHz and 1.39~GHz its dimensions are consistent with the restoring beam. We do not see the cross-shaped extensions reported by \citet{Wangetal19} from the short ($\sim$16~min) VLA 1.4~GHz observation, but we do see a small (8\arcs\ radius) extension to the southwest in our 1.39~GHz image. Our beam is smaller than that used in the VLA image (2.69$\times$2.29\arcs\ for GMRT 1.39~GHz compared to 5.6$\times$3.3\arcs\ for VLA 1.4~GHz) and the noise level of our image is low enough that the reported structures should be detected if they are present. At 333~MHz the beam is extremely elliptical, and there is a hint of extension to the northwest, but on scales considerably larger than those of the extensions reported by Wang et al. We will discuss results drawn from the radio data in Section~\ref{sec:radioims}.

\section{Results}
\label{sec:res}

Figure~\ref{fig:introims} shows \chandra\ X-ray and SDSS $i$-band images of the group. The 0.5-2~keV X-ray image combines all \chandra\ observations, smoothed with a 3\arcs\ Gaussian, and corrected with the combined exposure map. It shows the two cores and surrounding diffuse emission. This large-scale X-ray emission extends to the edge of the \xmms\ and ACIS-I fields of view ($>$13\arcm/410~kpc). The image shows a low surface brightness gap between the two cores, and that their surface brightness drops abruptly on the sides adjacent to this gap. These sharp discontinuities are the cold fronts identified by \citet{DupkeMartins13} and \citet{Wangetal19}. In the opposing direction both cores have drawn-out surface brightness distributions suggestive of tails. The southern tail is broad and relatively featureless. The northern tail appears narrower, curves to the northwest, and shows hints of bifurcation. The X-ray brightness peak in the southern core corresponds to the optical centroid of NGC~6338. The brightest X-ray emission in the northern core forms an east-west bar which overlaps the \Dtf\ ellipse of VII~Zw~700, but is located to the north of the centroid of the main galaxy.

To examine the structures in the X-ray halo in more detail, we fitted a simple surface brightness model, consisting of three $\beta$-models and a flat background component, all folded through the combined exposure map. Two of the $\beta$ models represent the cores. As these are asymmetric, the model components were fixed to be circular, and we therefore expect over-subtraction in the gap between the cores, where both models are likely to overestimate the emission. The third $\beta$-model represents the large-scale emission, and its ellipticity and position angle were fitted, as were all other parameters. We do not expect the fitted values to be physically meaningful. Our purpose in fitting the model was only to approximate the overall surface brightness distribution and then subtract it, to search for residual structures.

\begin{figure}
\includegraphics[width=\columnwidth,bb=36 126 577 667]{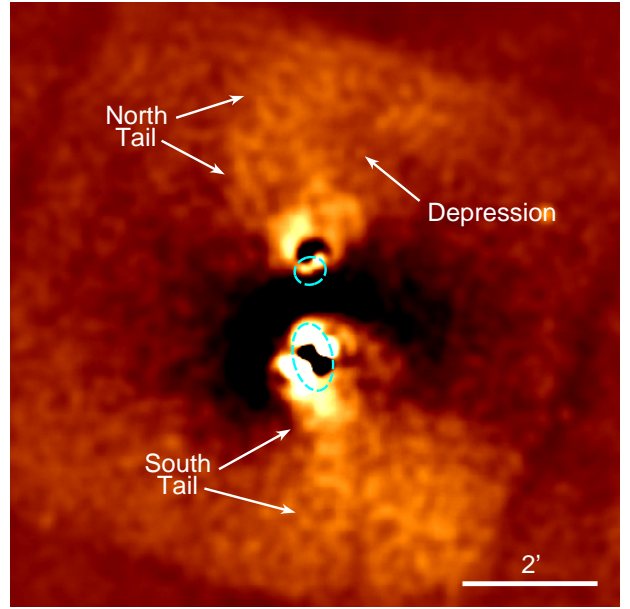}
\caption{\label{fig:tails}0.5-2~keV \chandra\ residual image after point source removal and refilling, and subtraction of a simple surface brightness model. The image has 1\arcs\ pixels, has been smoothed with a 10\arcs\ Gaussian, and uses a linear colour scale. Bright regions indicate excess emission above the model, the most notable features being the bright tails extending behind each core. The \Dtf\ ellipses of the dominant galaxies are marked by dashed lines. While the model includes an exposure map, some detector structures are still visible, e.g., the S3 CCD boundary at the edges of the field, and an ACIS-I chip gap as a vertical line running through the south tail.}
\end{figure}

Figure~\ref{fig:tails} shows the residual image, heavily smoothed (10\arcs\ Gaussian) to reveal large-scale features. Outside the cores, the strongest positive residuals are the tails, seen north of the north core and south-southwest of the south core. The eastern boundary of the northern tail appears to be fairly strongly curved to the northwest. Its western boundary is less distinct and includes a deep bay or depression behind the core. The strong negative residuals between the cores show the expected over-subtraction.

\begin{figure*}
\centerline{\includegraphics[width=0.9\textwidth,trim=50 290 70 290,clip]{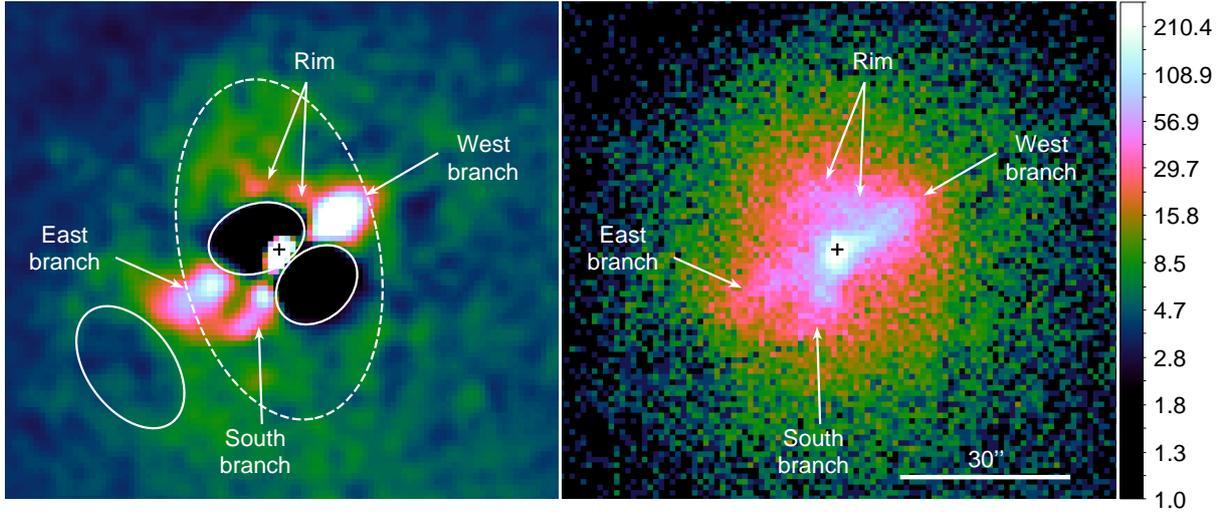}} 
\caption{\label{fig:Score} Combined \chandra\ 0.5-2~keV images of the south core. The \textit{right} panel shows a raw counts images with 1\arcs\ pixels, with structures labeled, and logarithmic colour scale indicated by the colour bar in units of counts per pixel. The \textit{left} panel shows a residual image, smoothed with a 3\arcs\ Gaussian, with a linear colour scale. Solid ellipses indicate possible cavities. The cross and dashed ellipse indicate the optical centroid and \Dtf\ contour of the dominant galaxy, NGC~6338.}
\end{figure*}

\begin{figure*}
\centerline{\includegraphics[width=0.9\textwidth,trim=50 290 70 290,clip]{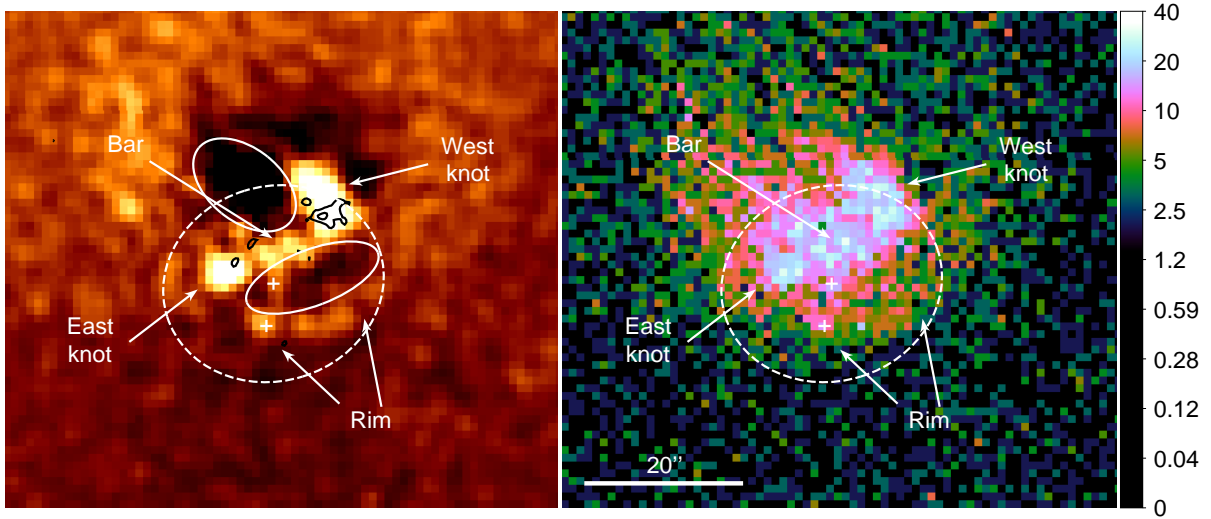}} 
\caption{\label{fig:Ncore} Combined \chandra\ 0.5-2~keV images of the north cores. Images are as described in Figure~\protect\ref{fig:Score}, except in that the \textit{left} panel is smoothed with a 2\arcs\ Gaussian, and crosses indicate the optical centroids of the two components of VII~Zw~700. APO H$\alpha$ contours are overlaid in black on the left panel.}
\end{figure*}

Figures~\ref{fig:Score} and \ref{fig:Ncore} show images of the two cores, both in raw \chandra\ 0.5-2~keV counts, and in the residuals to the surface brightness model. In the south core, the X-ray surface brightness peak is located at the optical centroid of NGC~6338. Bright X-ray emission extends across this peak on a northwest-southeast axis, splitting into two branches on the southeast side. The residual map shows dark regions of over-subtraction to the southwest and northeast of the X-ray peak; these were identified as cavities by \citet{Pandgeetal12}. Excess X-ray emission extends from the western branch around the northern edge of these negative residuals, forming an apparent rim.

The X-ray emission of the northern core is dominated by a bar of emission extending roughly southeast-northwest. The brightest part of this structure is at its western end, where a knot of emission extends perpendicular to the bar axis, to the northeast. Negative residuals are visible on either side of the bar, most notably to the northeast. Those to the southwest form a slight dip in brightness bounded by a roughly semi-circular rim of enhanced emission which seems to mark the edge of the diffuse emission associated with the core. As noted above, the optical centroids of the two components of VII~Zw~700 are not correlated with the brightest X-ray emission. The centroid of the larger component, MCG+10-24-117, falls just south of the bar while the smaller galaxy is centred in the rim. There is a hint of X-ray emission extending from the bar along the axis between the two galaxy centroids. The offset between the bar and optical centroids, and the position of the X-ray rim well inside the \Dtf\ ellipse of VII~Zw~700, suggest that the hot gas has been pushed back by external pressure and may now be at least partly disconnected from the stellar component of the dominant galaxy. We will examine these structures in more detail later.

\subsection{Spectral maps}
\label{sec:maps}

\begin{figure*}
\includegraphics[width=0.95\textwidth, bb=36 240 577 564]{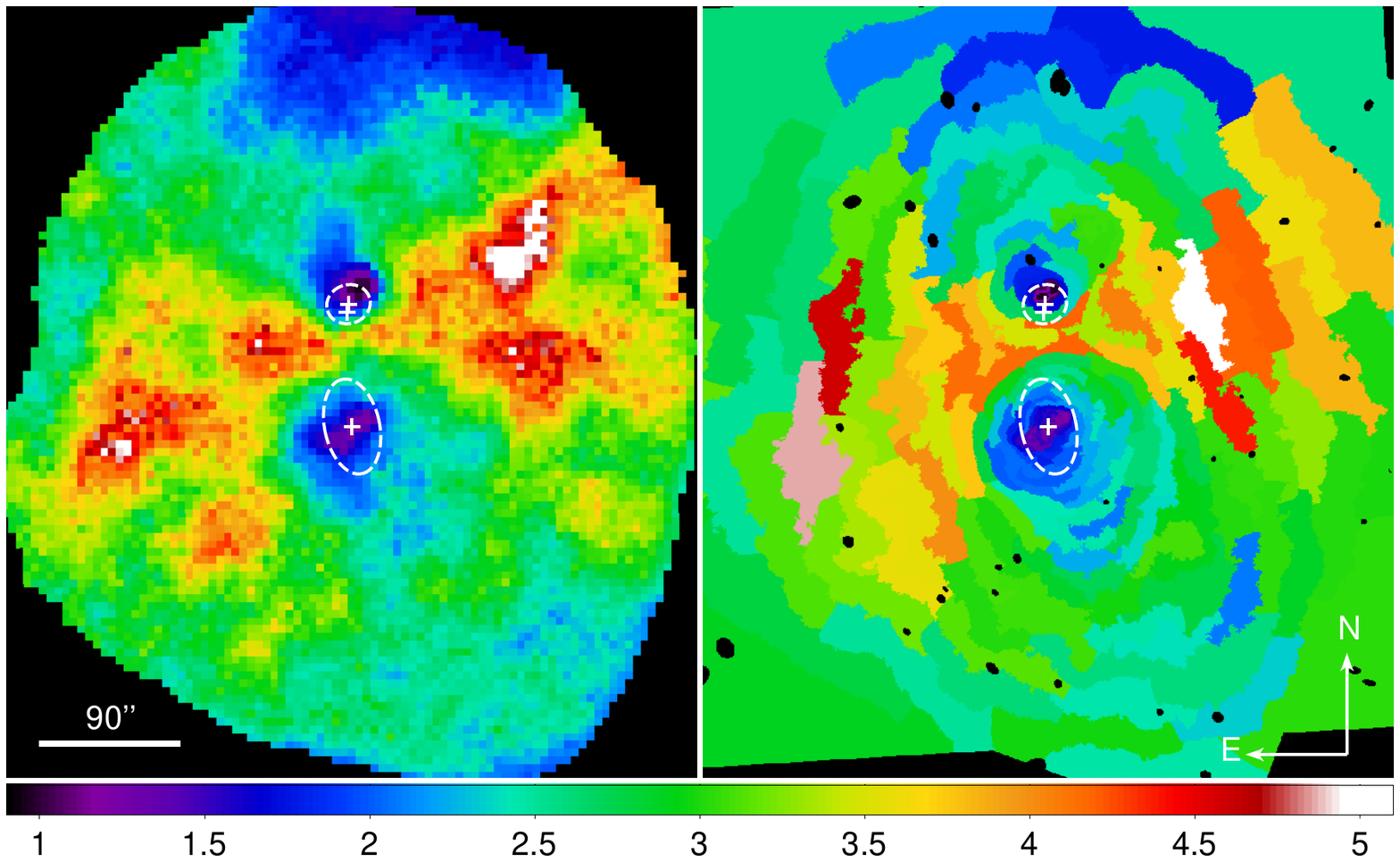}
\caption{\label{fig:kT}\chandra\ temperature maps, with matched colour scale in keV. The \textit{left} panel shows the 5\arcs-scale, 1500 net count ``fixed grid'' map, while the \textit{right} panel shows the S/N=30 contour binning map. Positions and SDSS \Dtf\ ellipses of the dominant galaxies are marked by crosses and dashed lines. }
\end{figure*}

Figure~\ref{fig:kT} shows \chandra\ temperature maps of the group, created using the fixed grid and contour binning methods. The maps clearly show cool (1-2~keV) temperatures associated with the dominant galaxies, and extending to the north and south along their tails. Higher temperatures (3.5-5~keV) are observed between the two cores, and to east and west. These high temperatures likely represent gas shock-heated by the merger. Outside the two cores, the coolest temperatures ($<$2~keV) are seen at the northern boundary of the map, and in general temperatures north of the northern core are somewhat cooler than those south of the southern core.

The two mapping methods generally agree quite well, but there are differences, most notably in the hottest regions. The fixed grid map suggests that west of the two cores the highest temperatures form a V-shaped pattern with cooler temperatures between its arms. This is not seen in the contour binning map; its regions follow the surface brightness, so are elongated north-south, overlapping the arms of the V, the region between, and sometimes the tails. They therefore show a range of temperatures, rather than the structure within the region. Conversely, the contour binning does well in tracing the curved boundaries of the two cores, the branching filamentary structure in the south core, and the bar and clumps in the north core.   

\begin{figure}
\includegraphics[width=\columnwidth,trim=40 110 40 110,clip]{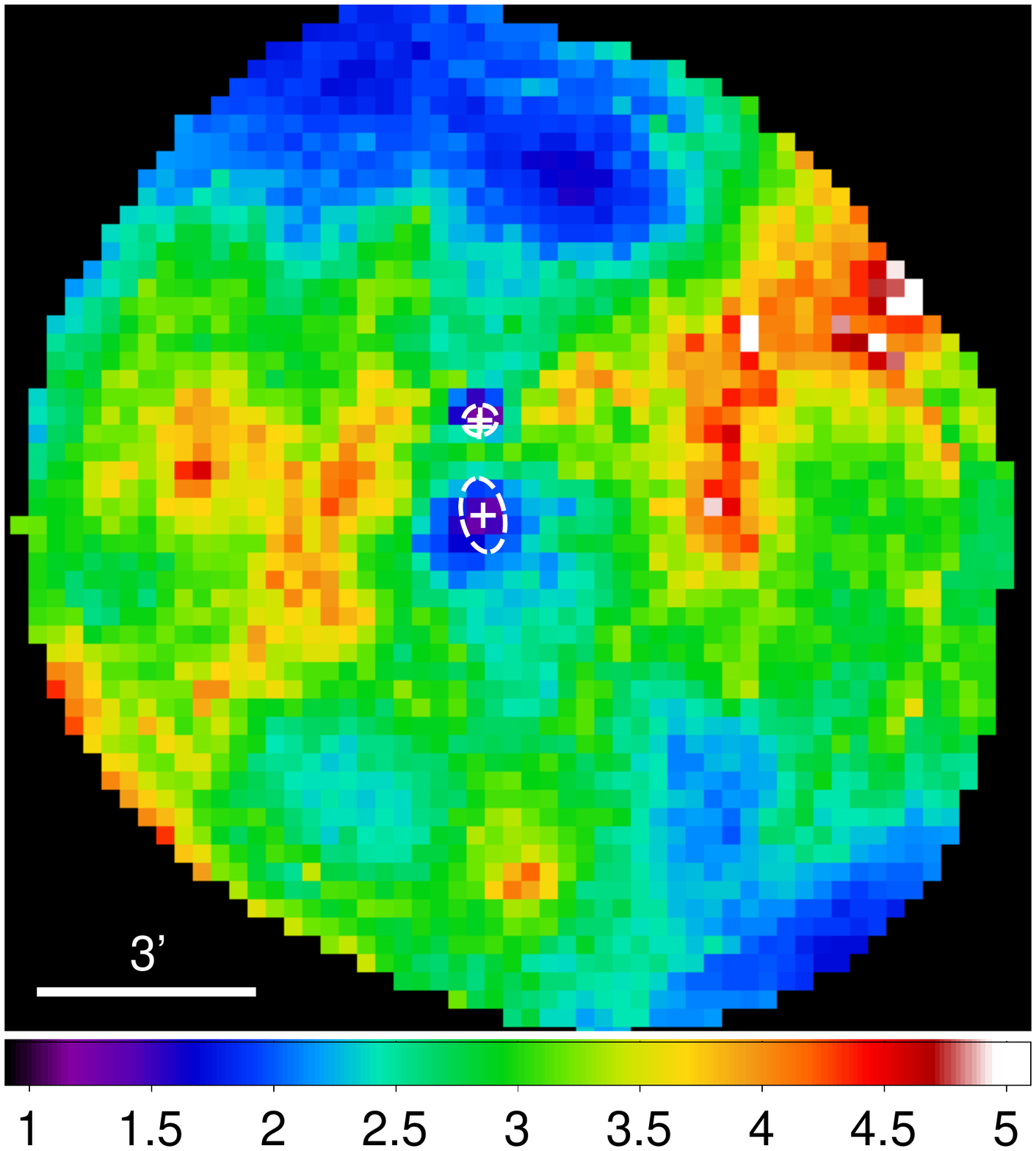}
\caption{\label{fig:XMMkT}\xmm\ 15\arcs-scale, 2000 net count temperature map, in units of keV. The colour scale and galaxy positions and \Dtf\ ellipses are matched to those in Figure~\ref{fig:kT}.}
\end{figure}

Figure~\ref{fig:XMMkT} shows the \xmms\ temperature map of the system, which has lower resolution (15\arcs\ pixels) but extends to larger radius than the \chandra\ maps (roughly 6\farcm8 radius compared to 4\arcm). The same general structure is seen, with cooler cores and tails, and the highest temperatures east and west of the cores. The agreement between \xmms\ and \chandra\ demonstrates the reliability of the approach.

\begin{figure}
\includegraphics[width=\columnwidth,trim=40 80 40 80,clip]{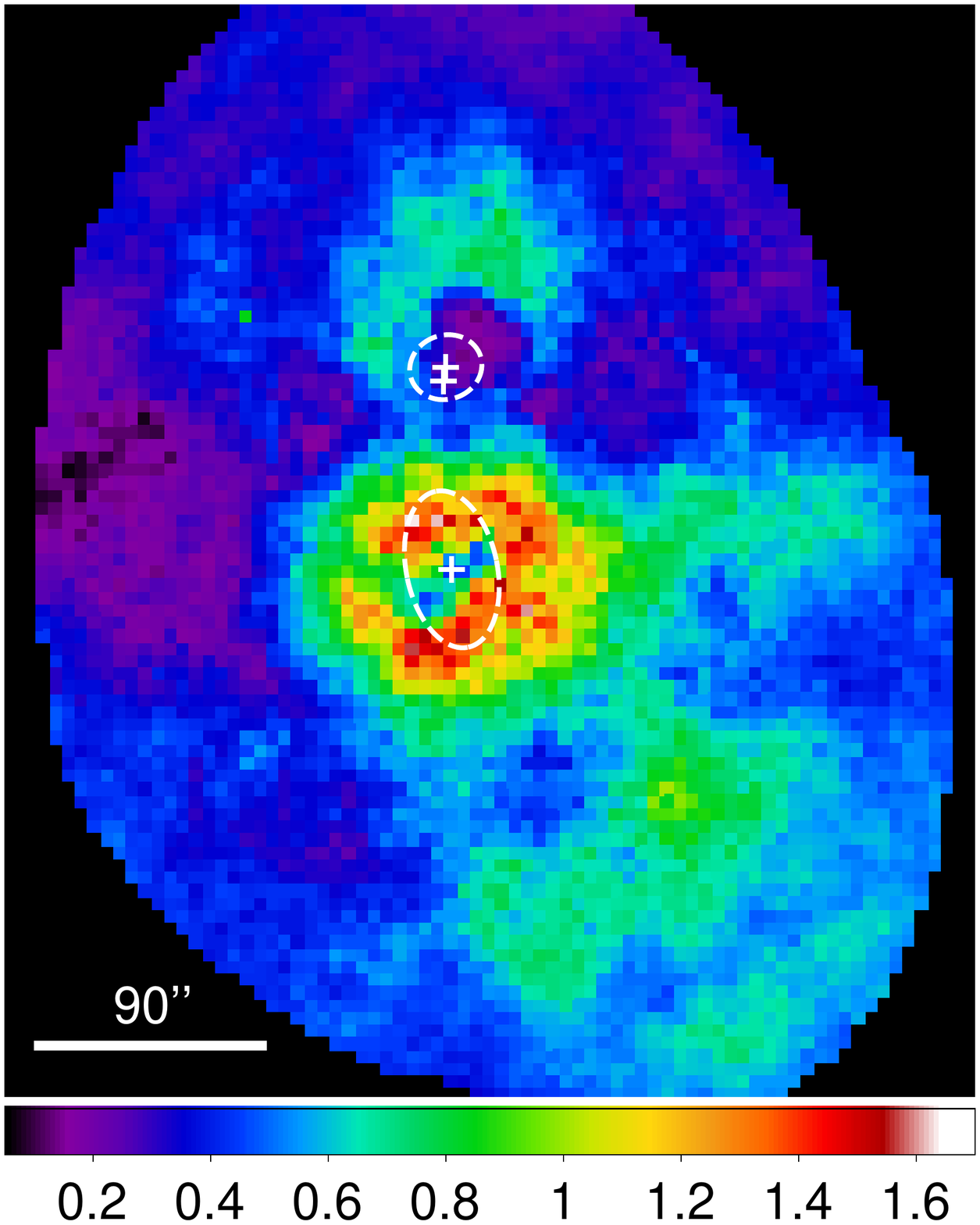}
\caption{\label{fig:Z}\chandra\ 5\arcs-scale, S/N=50 abundance map with scale in solar units. Positions and SDSS \Dtf\ ellipses of the dominant galaxies are marked by crosses and dashed lines. }
\end{figure}

Figure~\ref{fig:Z} shows the \chandra\ 5\arcs-scale, S/N=50 map of abundance. The more stringent S/N requirement is necessary for reliable abundance measurements, but means the map covers a slightly smaller area and is effectively more heavily smoothed. The south core is visible as a region of super-solar abundances surrounding NGC~6338, with the southern tail traced by regions of $\sim$0.7-0.8\Zsol\ abundances. The base of the northern tail is also visible as a region of $\sim$0.75\Zsol\ abundances, but the northern core shows low abundances. This is caused by the Fe-bias effect \citep{Buote00b}; fitting a single-temperature model to multi-temperature plasma emission at temperatures $\sim$1~keV results in an abundance biased to low values. A similar, though smaller, central dip in abundances can be seen in the south core. Outside the cores and tails, the abundance is typically $\sim$0.2-0.4\Zsol.

\begin{figure*}
\includegraphics[width=0.49\textwidth,trim=40 85 40 100,clip]{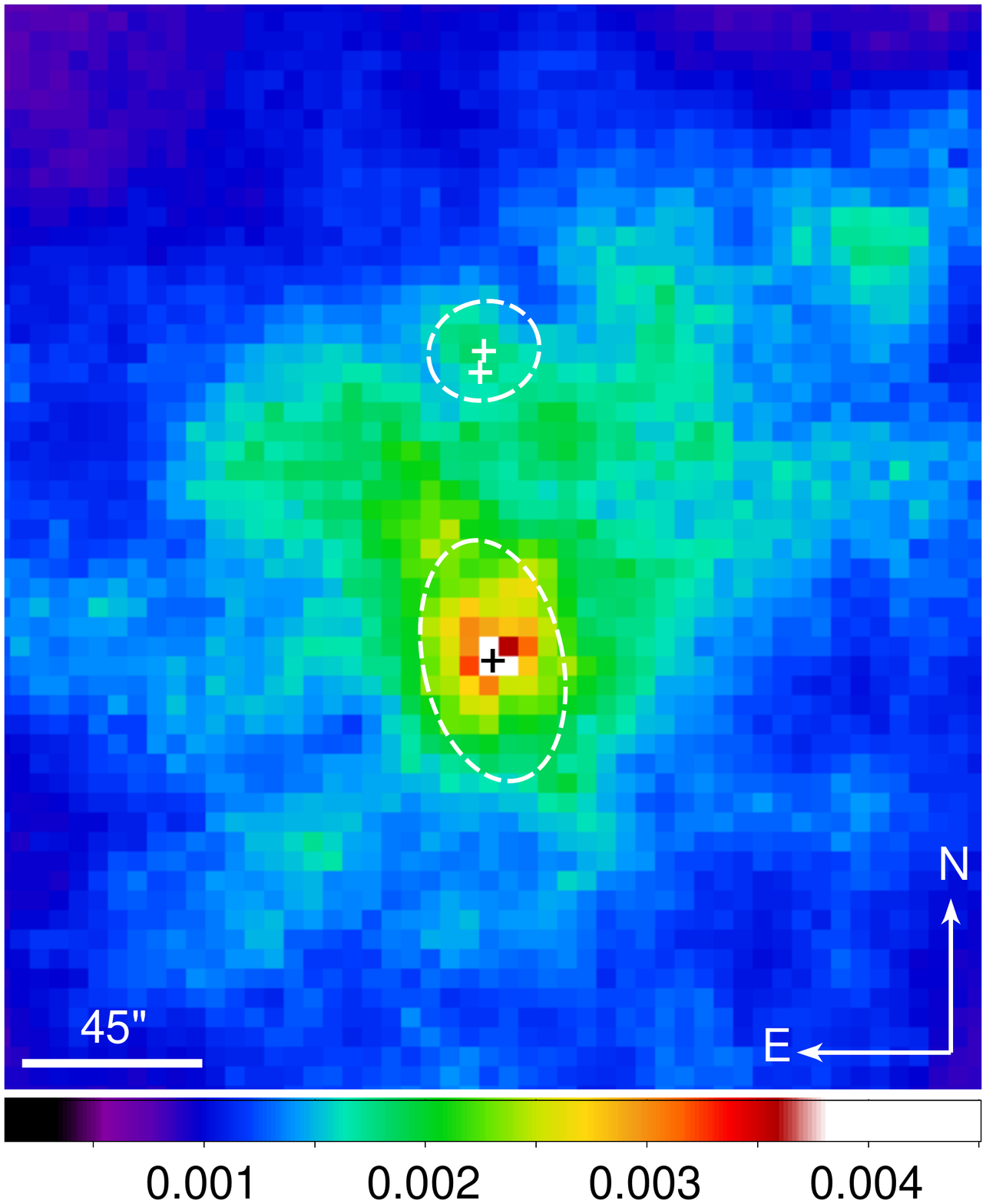}
\includegraphics[width=0.49\textwidth,trim=40 85 40 100,clip]{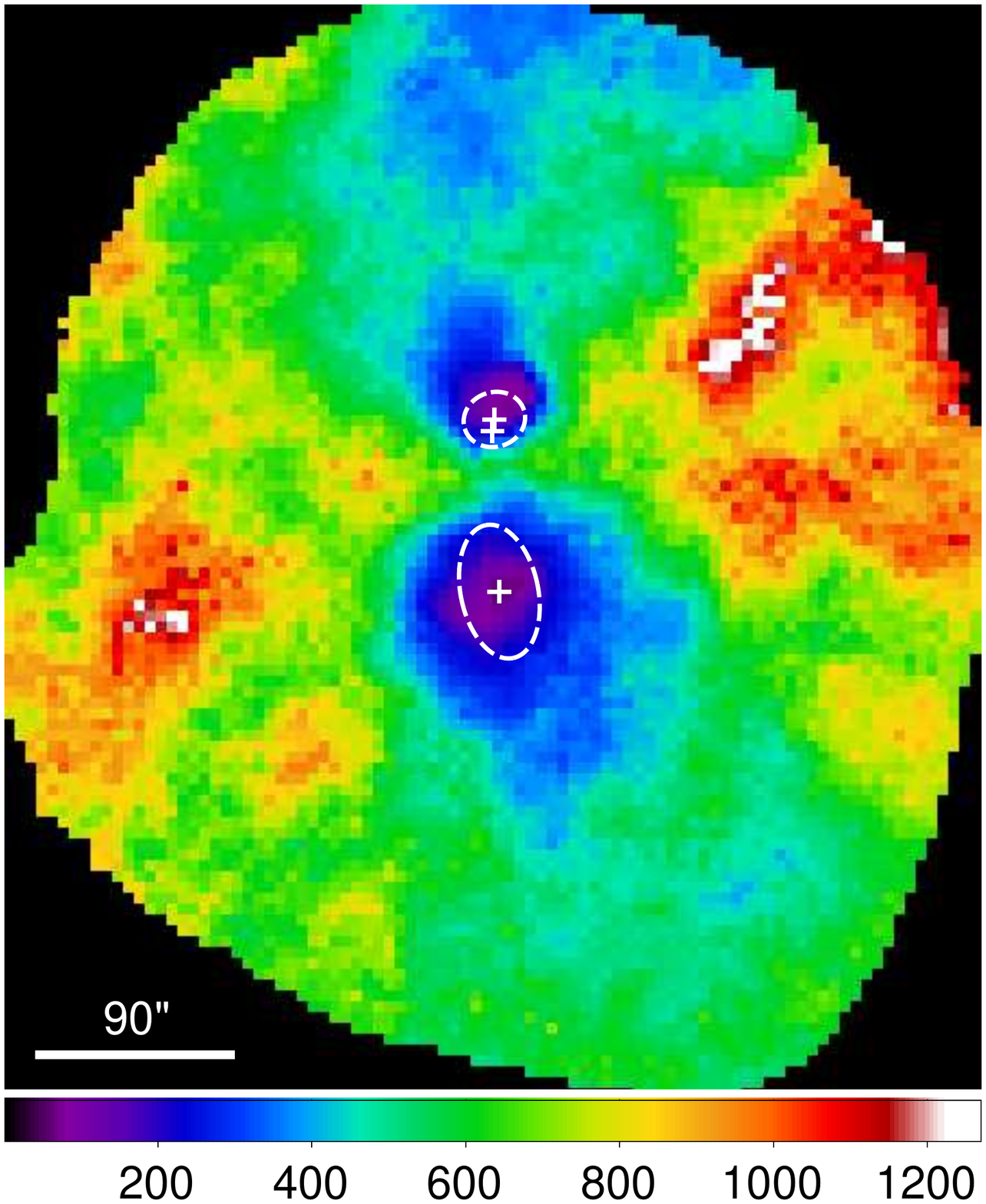}
\caption{\label{fig:PS}\chandra\ 5\arcs-scale, 1500 net count pseudo-pressure (\textit{left}) and pseudo-entropy maps. Units are arbitrary and do not indicate the true entropy or pressure. Positions and SDSS \Dtf\ ellipses of the dominant galaxies are marked by crosses and dashed lines. }
\end{figure*}

Figure~\ref{fig:PS} shows a \chandra\ pseudo-pressure map of the region around the two cores, and a pseudo-entropy map of the system as a whole. The pseudo-entropy map shows features similar to those in the temperature maps shown in Figure~\ref{fig:kT} but in some regards shows the gas structures more clearly. The lower entropies of the cores and tails show up in blue and green, while the hotter, high-entropy gas is in red and yellow. The V-shaped high temperature structure west of the cores is visible, and both tails appear to trend to the west of the north-south axis of the cores, making the western high-entropy region narrower. The pseudo-pressure map suggests that while the highest pressures are found in the core of NGC~6338, high pressures extend north into the gap between the cores. The high pressure area is widest in this gap, extending to east and west of the two dominant galaxies. This suggests that the region between the two cores contains material strongly shock-heated and raised to high pressures by the merger. VII~Zw~700 is correlated with a small area of high pressure, but seems comparable with pressures immediately to its south. The pseudo-pressure map does not show any structures at the position of the surface brightness discontinuities, as expected if they are cold fronts.

\citet{Laganaetal19} have recently published \xmms\ spectral maps of NGC~6338. Their maps, made using a technique similar to our ``fixed grid'' method, with 25\arcs\ resolution and a minimum of 1500 net counts per spectral region, are in good agreement with ours. The only discrepancy appears in their abundance map, where a region of enhanced abundance is visible east of NGC~6338. This is not visible in our deep \chandra\ data, nor in our \xmms\ abundance map, and may be the product of fitting uncertainties associated with the high temperatures in this area.

\subsection{Radial gas profiles}
\label{sec:global}

\begin{figure*}
\includegraphics[width=\textwidth,bb=20 220 570 779]{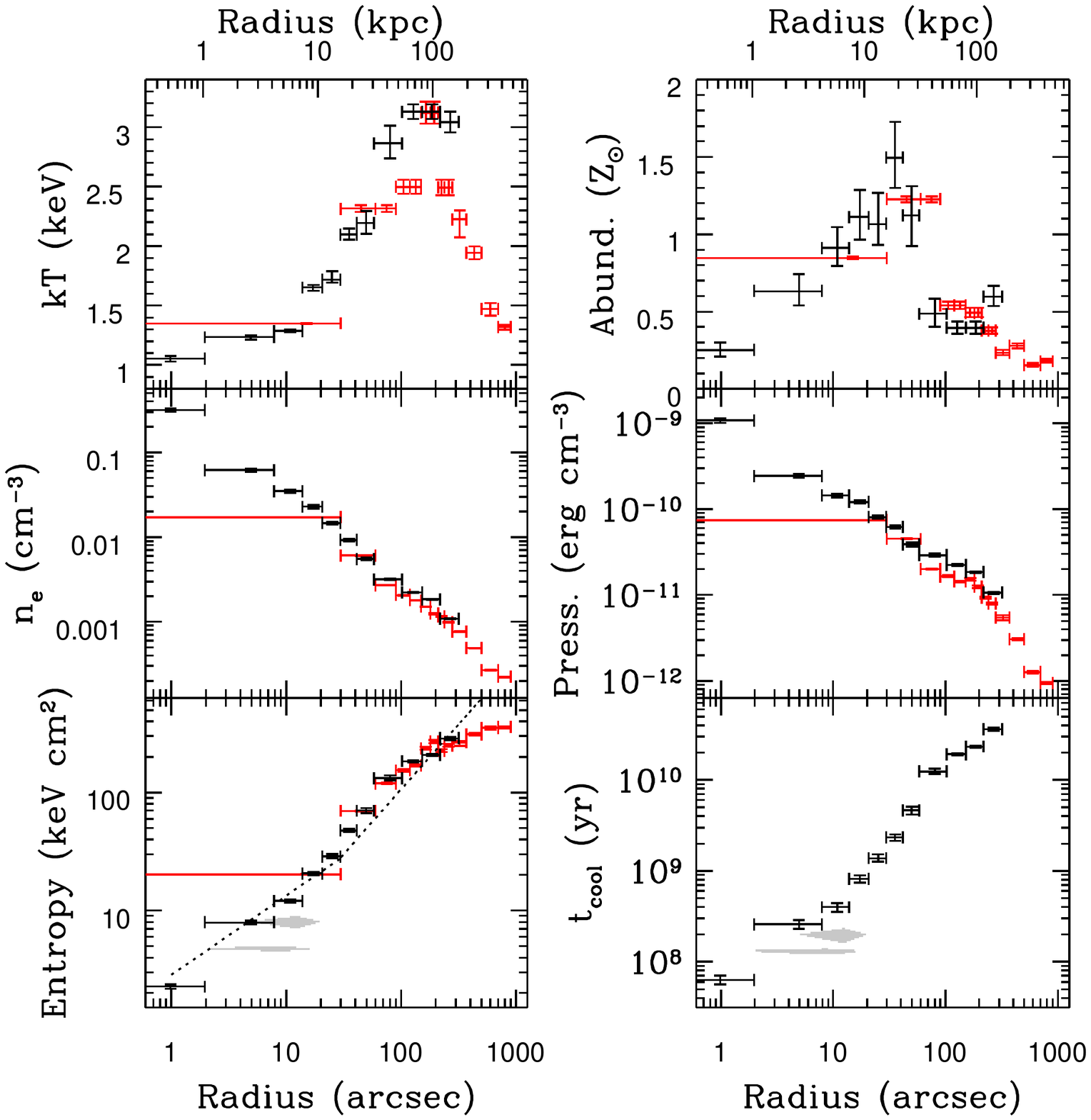}
\caption{\label{fig:profiles} Deprojected radial profiles of temperature, abundance, density, pressure, entropy and (for \chandra) isochoric cooling time, based on annuli centred on NGC~6338, and excluding VII~Zw~700 and its tail. Black points represent \chandra\ and red points \xmms. Grey diamonds indicate cooling time and entropy of the three X-ray filaments in the south core. The dotted line in the entropy plot shows a relation with slope r$^{0.67}$ within 30\arcs\ and r$^{1.1}$ at larger radii, roughly scaled to match the inner data points.}
\end{figure*}

Figure~\ref{fig:profiles} shows deprojected profiles of temperature, abundance, density, pressure, entropy and cooling time for the system. These were extracted using annuli centred on the optical centroid of NGC~6338, and excluding the north core and tail. In such a disturbed and asymmetric system, the assumptions implicit in spherical deprojection are invalid, but it should still produce reasonably reliable results in the (relatively circular) south core. 

The \chandra\ spectra were extracted from regions selected to have signal-to-noise ratios of $\geq$125. \xmms\ spectra were extracted from regions with S/N$\geq$50, with lower values at large radii, where additional annuli were included to help constrain the background model. Deprojection was performed using the \textsc{projct} model in \textsc{Xspec}. Temperature and abundance were tied in a few pairs of bins to prevent the "ringing" effect and reduce uncertainties. In general the profiles derived from the two satellites are in reasonable agreement, taking into account their different responses and the effects of averaging regions of different temperature and abundance within annuli. Entropy is defined as $kTn_e^{-2/3}$ where $kT$ is the X-ray temperature in keV and $n_e$ the electron number density in cm$^{-3}$. Pressure is defined as 2$kTn_e$, and the isochoric cooling time as:

\begin{equation}
t_{cool} = 5.076\times10^{-17}\times\frac{3kTn_eV\mu_e}{2\mu L_X},
\end{equation}

\noindent where the units of $t_{cool}$ are years, $V$ is the volume of the gas in cm$^3$, $L_X$ its bolometric luminosity in erg~s$^{-1}$, and $\mu$ and $\mu_e$ are the mean molecular weight (0.593) and the mean mass per electron (1.167) respectively. Note that we have calculated the isochoric cooling time to allow easy comparison with cooling threshold established in prior studies. The isobaric cooling time is greater by a factor 5/3 and allows for work done on each shell of gas as it cools at constant pressure.

The temperature profile shows a peak of $\sim$3~keV at a radius of $\sim$100~kpc ($\sim$190\arcs), with a $\sim$1~keV core and temperature declining to $\sim$1.3~keV at $\sim$420~kpc ($\sim$800\arcs). Abundance peaks at $\sim$20~kpc ($\sim$40\arcs), with a low-abundance core probably indicating the Fe-bias effect, as in the abundance map. Core entropy and cooling time are both extremely low, 2.26$\pm$0.09~keV~cm$^2$ and 63$\pm$7~Myr, and cooling time is below 1~Gyr within $\sim$10.9~kpc (20.6\arcs). Comparison with the mean central entropy profile (with slope $\propto{\rm r}^{0.67}$) found by \citet{Panagouliaetal14} shows a good match with our data at radii $<$16~kpc. At larger radii the entropy profile steepens and then flattens again, and is a poor match to the r$^{1.1}$ profiles found in most groups and clusters \citep{Voitetal05}.

We also extracted profiles from the north core and tail, the south core and tail, and a pair of sectors extending east and west excluding both cores and tails. We find results in good agreement with the spectral maps, with higher abundances in the cores and tails, and higher temperatures in the east-west region. For both the east-west region and the southern core and tail, projected temperatures fall below 2~keV at about 4.5\arcm\ ($\sim$140~kpc) from NGC~6338, and reach $\sim$1.3~keV beyond $\sim$7.5\arcm ($\sim$240~kpc). The northern core and tail is cooler than its surroundings at all radii, and its temperature falls below 2~keV $\sim$2.4\arcm\ (75~kpc) from the centroid of VII~Zw~700. In the north core, cooling times within the central 12\arcs\ ($\sim$6.5~kpc, the brightest part of the core) are $\sim$1-5~Gyr, and entropies are 20-40\kevcmsq.

\subsection{Multiphase gas in the cores}
\label{sec:multiphase}

\begin{figure*}
\includegraphics[width=0.49\textwidth,bb=36 95 577 697]{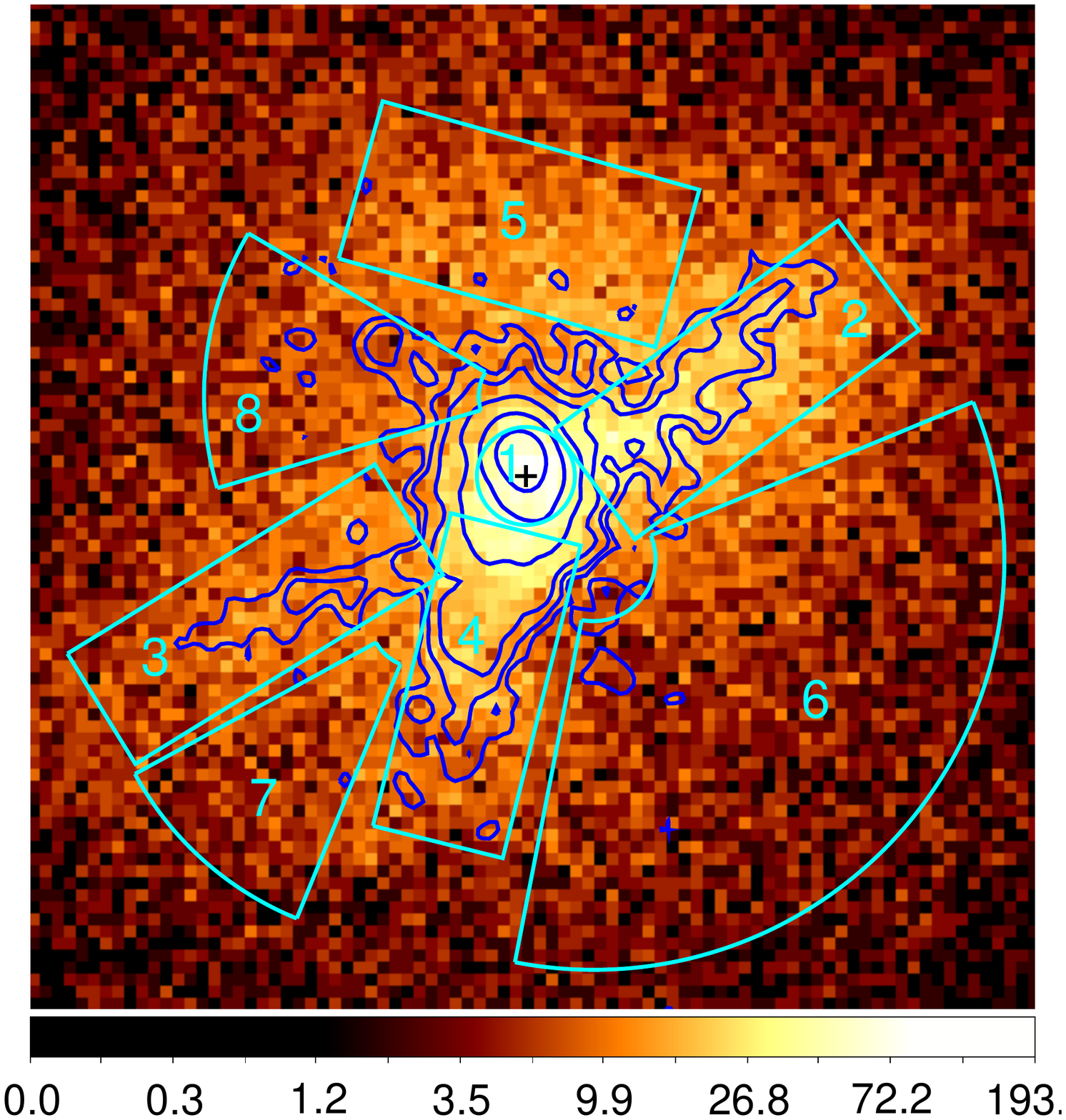}
\includegraphics[width=0.49\textwidth,bb=36 95 577 697]{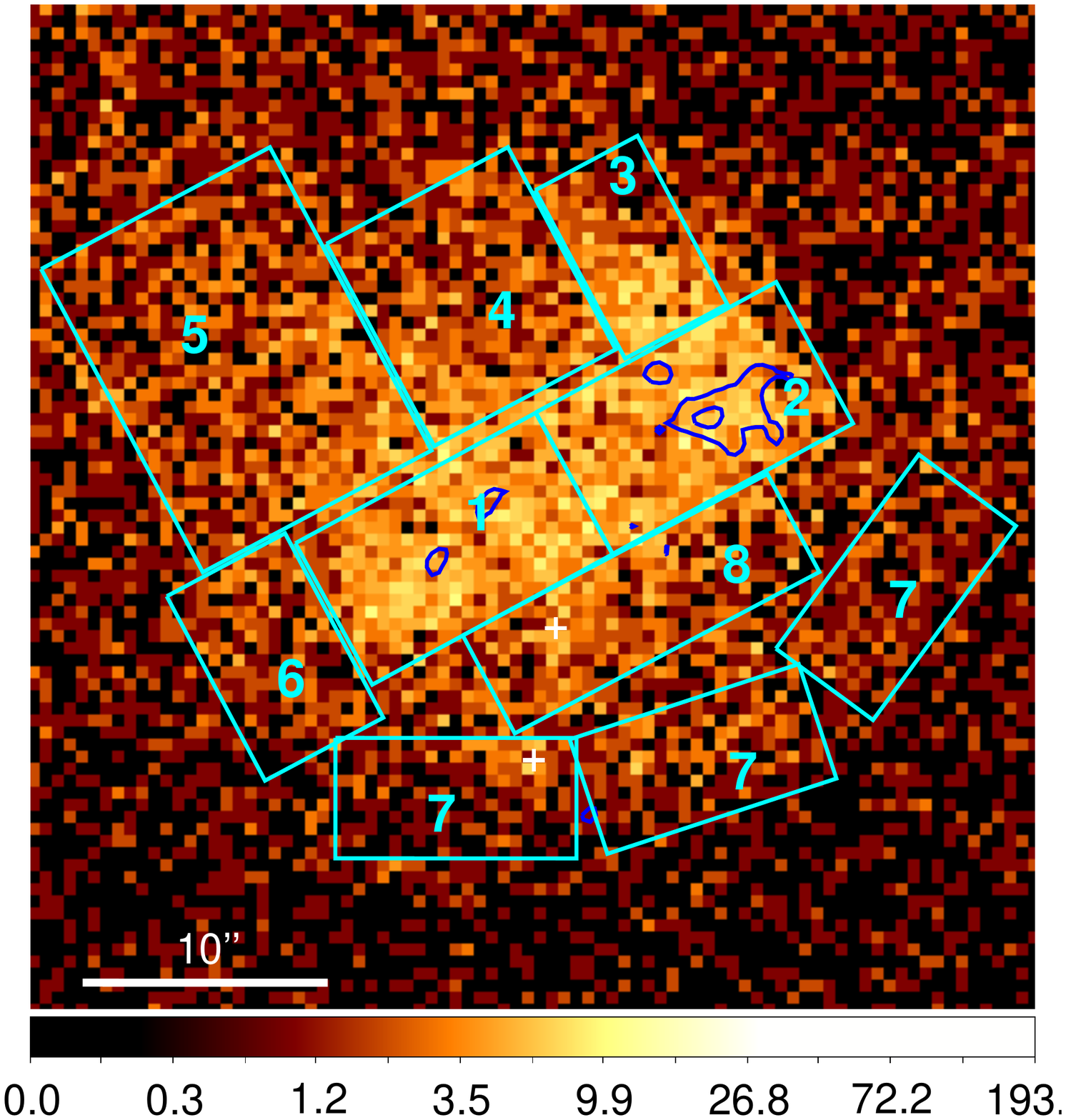}
\caption{\label{fig:cores} \chandra\ 0.5-2~keV images of the south (\textit{left}) and north (\textit{right}) cores in units of counts per pixel, with APO H$\alpha$ contours overlaid. Optical centroids for the dominant galaxies are marked by crosses, and regions used for spectral extraction are marked and numbered in cyan. The images have the same angular scale.} 
\end{figure*}

Figure~\ref{fig:cores} shows the two cores, overlaid with H$\alpha$ contours from the APO imaging (see also Figure~\ref{fig:Ncore}). In the south core, our H$\alpha$ imaging detects the three emission-line filaments seen by
\citet{Marteletal04} and \citet{Gomesetal16b}, tracing them out to
$\sim$17\arcs\ from the nucleus of NGC~6338. The H$\alpha$ emission is strongly correlated with the X-ray structure, with H$\alpha$ filaments extending along the east, west and south branches of the X-ray ridge shown in Figure~\ref{fig:Score}. The H$\alpha$ peak also coincides with the optical and X-ray peaks. However the H$\alpha$ also extends to the northeast and southwest, into the regions of low surface brightness which \citet{Pandgeetal12} identified as cavities. 

The total H$\alpha$ flux from the south core is $\sim$
1.95$\times$10$^{-14}$\ergpspcmsq\, assuming these ratios between the [NII]
lines and the H$\alpha$ line: [NII]6583/H$\alpha$ = 2 and [NII]6548/H$\alpha$ =
0.67 \citep[from][and our DIS long-slit spectrum]{Gomesetal16b}.  The total
H$\alpha$ luminosity is 3.1$\pm$0.6$\times$10$^{40}$\ergps.  With the
distance used by Gomes et al., the total H$\alpha$ luminosity is
3.2$\times$10$^{40}$\ergps, which is consistent within uncertainties with
their luminosity estimate from the integral field spectroscopy data,
2.94$\times$10$^{40}$\ergps.

VII~Zw~700 was not covered by CALIFA. The SDSS spectrum reveals a
relatively strong [NII] 6584 line, H$\alpha$ absorption and a weak [NII] 6548 line
in the nuclear region.  Interestingly, our narrow-band imaging data reveal
an emission-line blob $\sim$11\arcs\ northwest of VII~Zw~700, coincident
with the bright X-ray knot at the western end of the bar (see
Figure~\ref{fig:Ncore}). The position of the denser
  H$\alpha$-emitting ionized gas within this knot suggests that it is
  shielded from any external pressures associated with the merger,
  otherwise the denser gas would lead the more diffuse X-ray-emitting
  component. The DIS long-slit observation ran across this blob, and
confirms that it is an emission-line object with H$\alpha$, H$\beta$, [OI]
6300, two [NII] lines and two [SII] lines detected. Its heliocentric
velocity is 9665$\pm$27\kmps, confirming its association with
VII~Zw~700. The [NII] 6584 / H$\alpha$ ratio is $\sim$0.8, which is too
high for HII regions so the ionization source is not young stars.  This
ratio is typical of many emission-line nebulae in cool cores
\citep[e.g.,][]{Tremblayetal18}.  With the line ratios from the DIS data
([NII]6583/H$\alpha$ = 0.8 and [NII]6548/[NII]6583 = 1/3), the total
H$\alpha$ luminosity of this blob is 3.1$\times$10$^{39}$\ergps.

To determine the state of the hot X-ray emitting gas in the core
structures, we extracted spectra from the numbered regions shown in
Figure~\ref{fig:cores}. In each region, we attempted to fit one and
two-component APEC thermal plasma models, and determined whether the
two-component model provided a better fit at greater than 98 per cent
(estimated from the reduced $\chi^2$ values of each fit using the
\textsc{ftest} task in \textsc{Xspec}). Abundance was left free in the fits
where possible, and tied between components in the two-temperature fits.
The results are shown in Table~\ref{tab:regfits}.

\begin{table*}
\caption{\label{tab:regfits} Best fitting temperatures, abundances, and normalizations for one and two-temperature APEC fits to the regions shown in Figure~\protect\ref{fig:cores}, with the reduced $\chi^2$ and number of degrees of freedom of the fit. Abundance was unconstrained for region 6 in the north core, so was held fixed at the mean value for the core.}
\begin{center}
\begin{tabular}{llccccc}
\hline
Region & notes & kT$_{cool}$ & kT$_{hot}$ & Abund. & N$_{cool}$/N$_{hot}$ & red. $\chi^2$/d.o.f.  \\
       &       & (keV)  & (keV)  & (\Zsol) & & \\
\hline
\multicolumn{6}{l}{\textit{South core}} \\
1 & core     & 1.01$_{-0.04}^{+0.03}$ & 1.79$_{-0.18}^{+0.20}$ & 0.77$_{-0.18}^{+0.25}$ & 0.96$^{+0.35}_{-0.31}$ & 0.859/100 \\
2 & W branch & 1.08$_{-0.02}^{+0.04}$ & 1.72$_{-0.08}^{+0.06}$ & 0.89$_{-0.10}^{+0.22}$ & 1.04$^{+0.23}_{-0.27}$ & 1.216/149 \\
3 & E branch & 1.25$_{-0.17}^{+0.07}$ & 1.93$_{-0.28}^{+0.64}$ & 1.27$_{-0.30}^{+0.42}$ & 0.77$^{+0.74}_{-0.64}$ & 1.493/107 \\
4 & S branch & 1.05$_{-0.04}^{+0.03}$ & 1.57$_{-0.11}^{+0.13}$ & 0.99$_{-0.19}^{+0.24}$ & 0.54$^{+0.26}_{-0.19}$ & 1.017/137 \\
5 & rim      & -                      & 1.61$\pm$0.02          & 1.27$_{-0.16}^{+0.18}$ & - & 1.185/126 \\
6 & S cavity & -                      & 1.77$\pm$0.04          & 1.18$_{-0.11}^{+0.13}$ & - & 1.241/183 \\
7 &          & 0.88$_{-0.06}^{+0.13}$ & 1.74$_{-0.07}^{+0.16}$ & 1.76$_{-0.50}^{+0.98}$ & 0.11$^{+0.08}_{-0.05}$ & 1.247/62 \\
8 & N cavity & 1.32$_{-0.15}^{+0.06}$ & 2.38$_{-0.37}^{+0.39}$ & 2.67$_{-0.78}^{+1.27}$ & 0.41$^{+0.33}_{-0.25}$ & 1.045/87 \\
\multicolumn{6}{l}{\textit{North core}} \\
1 & bar, E knot & 0.83$_{-0.08}^{+0.18}$ & 1.52$_{-0.13}^{+0.46}$ & 0.60$_{-0.18}^{+0.27}$ & 0.36$^{+0.38}_{-0.12}$ & 1.257/52 \\
2 & bar, W knot & -                      & 0.94$_{-0.03}^{+0.03}$ & 0.22$_{-0.04}^{+0.05}$ & - & 0.984/52 \\
3 & tail west   & -                      & 0.99$_{-0.05}^{+0.05}$ & 0.22$_{-0.06}^{+0.09}$ & - & 0.962/18 \\
4 & tail centre & -                      & 1.52$_{-0.08}^{+0.07}$ & 0.67$_{-0.19}^{+0.26}$ & - & 0.598/32 \\
5 & tail east   & -                      & 1.64$_{-0.04}^{+0.04}$ & 0.84$_{-0.17}^{+0.21}$ & - & 1.206/56 \\
6 &             & -                      & 2.06$_{-0.17}^{+0.22}$ & 0.60                   & - & 1.622/14 \\
7 & rim         & -                      & 2.03$_{-0.14}^{+0.13}$ & 0.59$_{-0.20}^{+0.27}$ & - & 0.948/39 \\
8 & cavity      & 0.87$\pm$0.13          & 2.25$_{-0.32}^{+0.49}$ & 0.57$_{-0.28}^{+0.54}$ & 0.19$^{+0.24}_{-0.09}$ & 0.896/28 \\
\hline
\end{tabular}
\end{center}
\end{table*}

Two-temperature models provide superior fits for most of the south core, the exceptions being regions 5 and 6, the rim and south cavity. Region 8, the north cavity, has the highest temperatures in both components, and a high, fairly poorly constrained abundance. This may indicate hotter material along the line of sight, or that the cool component is relatively weaker in this region than elsewhere in the south core. Region 1, the central 2\arcs-radius region coincident with the optical and H$\alpha$ peaks of NGC~6338, has the lowest temperature and abundance in the south core. We tested whether an additional model component was necessary, or whether an APEC+powerlaw model gave a better fit. The two-temperature model was clearly favoured over the APEC+powerlaw model.  

Regions 2, 3 and 4 cover the three filaments. For these regions, we also try fits using the spectrum of region 6 as a local background. A single APEC model provides a good fit in each case, and we find temperatures of 1.10$\pm$0.02, 1.18$^{+0.04}_{-0.05}$ and 1.26$^{+0.01}_{-0.02}$~keV for the W, E and S branches respectively. Abundances are lower than those found from the two-temperature fits, 0.57$^{+0.11}_{-0.09}$, 1.18$^{+0.61}_{-0.33}$ and 0.56$^{+0.07}_{-0.06}$\Zsol\ respectively. Approximating the filaments as cylinders, we can estimate their entropy and cooling time. The W and S branches have similar entropies (4.8$\pm$0.2 and 4.8$\pm$0.1\kevcmsq) and cooling times (129$\pm$8 and 134$\pm$5~Myr). The E branch (which is narrower in both X-ray and H$\alpha$, and at slightly larger radii from the nucleus) has higher values, 8.0$\pm$0.9\kevcmsq\ and 199$^{+37}_{-33}$~Myr. These values are included on Figure~\ref{fig:profiles}, which shows that they fall below the overall radial profile for the core, as expected.

We find less evidence of multi-temperature gas in the north core, with only regions 1 and 8 showing evidence of a second component. However, the low temperatures and abundances found in the single-temperature fits to regions 2 and 3 (the western knot and its tail) suggest these regions are also dominated by cool gas, with the Fe-bias effect reducing the apparent metallicity. Region 2 contains the offset H$\alpha$ cloud, supporting the possibility that this is a region of active or recent cooling. The remaining regions have temperatures of 1.5-2~keV, with the highest single temperatures found at the eastern end of the bar and in the rim, both regions just inside the surface brightness edge bounding the core.

\subsection{Active Nuclei}
\label{sec:Xagn}
As noted in section~\ref{sec:multiphase}, the X-ray spectrum of the central 2\arcs\ of NGC~6338 is best fitted by a two-temperature thermal plasma model. If a powerlaw is added to the two-temperature model, its index is not constrained, and if it is fixed at $\Gamma$=1.65 or 2.0, the model normalisation falls to zero. We find 3$\sigma$ upper limits on the AGN 2-10~keV X-ray flux of 8.61$\times$10$^{-15}$\ergpspcmsq\ for $\Gamma$=1.65 and 8.54$\times$10$^{-15}$\ergpspcmsq\ for $\Gamma$=2.0, equivalent to an upper limit on luminosity of L$_{2-10~keV}$$\leq$1.35$\times$10$^{40}$\ergps.

We therefore agree with \citet{Torresietal18}, who concluded that the AGN of NGC~6338 is X-ray faint, and place stronger limits on its luminosity. It should be noted that the high central density found in NGC~6338 (see Section~\ref{sec:global}) is therefore very unlikely to be biased by AGN emission.

We performed a similar analysis for the two galaxies of VII~Zw~700, using 2\arcs\ radius circular regions centred on their optical nuclei. As mentioned previously, imaging shows excess X-ray emission at the position of the two nuclei, with a narrow extension linking them and the bar structure. For the dominant member of the galaxy pair, MCG+10-24-117, we find that an APEC+powerlaw spectrum provides the best fit, though it is only superior to a simple APEC model at $\sim$95\% significance. The small number of counts forces us to fix abundance at 0.6\Zsol\ and the power-law index to $\Gamma$=1.65. With these parameters, we find the luminosity of the powerlaw component to be L$_{2-10~keV}$=4.17$^{+1.64}_{-1.68}$$\times$10$^{39}$\ergps. A two temperature APEC model does not provide an acceptable fit, as the temperature of the second component becomes unphysically large. For the smaller galaxy, 2MASS~J17152337+5725530, we find that an APEC model alone provides an acceptable fit. If we include a $\Gamma$=1.65 powerlaw, the 3$\sigma$ upper limit on its luminosity is L$_{2-10~keV}$$\leq$2.50$\times$10$^{38}$\ergps. 

\subsubsection{Radio imaging and spectrum}
\label{sec:radioims}

\begin{figure}
\includegraphics[width=\columnwidth,bb=36 149 577 644]{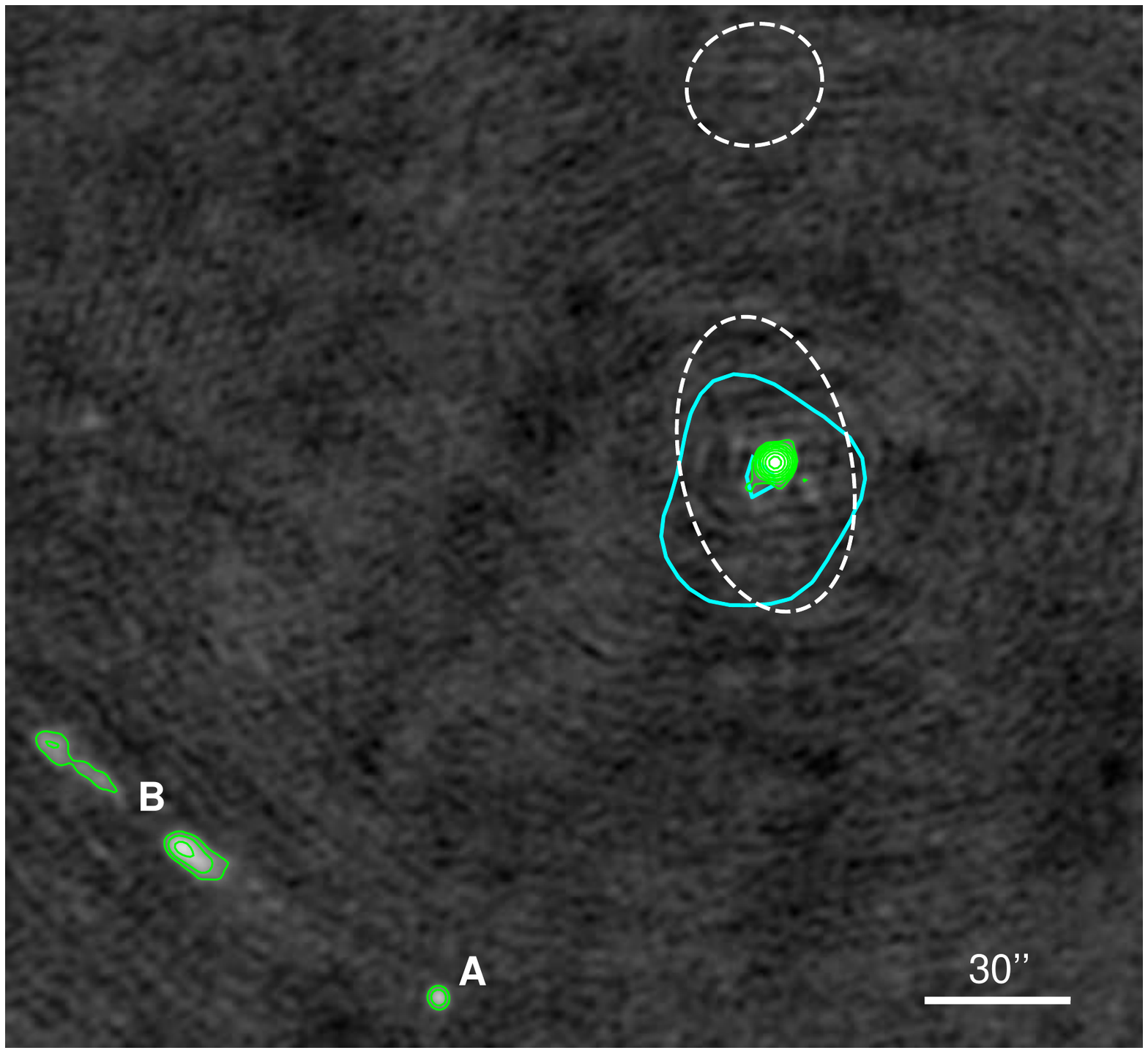}
\caption{\label{fig:radio}1.39~GHz GMRT image of NGC~6338, made with a 3\arcs\ HPBW circular restoring beam. \Dtf\ ellipses for the two BGGs are marked by dashed lines. Thin green contours indicate significant 1.39~GHz emission, starting at 5$\times$rms and increasing in steps of factor 2. Thicker cyan contours indicate 147~MHz emission, starting at 4$\times$rms and increasing by factors of 2. Source A corresponds to the $z$=0.065 galaxy 2MASS~J17153152+5722504, while the two components of source B corresponds to NVSS~J171538+572322.}
\end{figure}

Figure~\ref{fig:radio} shows the 1.39~GHz GMRT image of the group, made with a 3\arcs\ Half-power beam width (HPBW) restoring beam. We detect the radio source associated with the nucleus of NGC~6338, but see no radio emission from VII~Zw~700. As mentioned in Section~\ref{sec:radio}, there is a hint of extension to the southwest in the 5$\sigma$ contour. If this is real, it would be roughly aligned with the filamentary structures. For comparison, we have also overlaid the 147~MHz contours; at this lower resolution, the source is unresolved.

\begin{table}
\caption{\label{tab:radflux} Radio flux density measurements for NGC~6338, from our own analysis and the literature. Uncertainties on our GMRT measurements include a calibration uncertainty of 10\% at 147 and 333~MHz, and 8\% at 1388~MHz \citep{Chandraetal04a}. The third column lists the Half Power Beam Width (HPBW) for the restoring beam. Literature values are taken from: the VLA low-frequency sky survey redux \citep[VLSSr,][]{Laneetal14}; WENSS \citep{Rengelinketal97}; NVSS \citep{Condonetal98}; \citet{Beckeretal91}; and \citet{Marchaetal01}. }
\begin{center}
\begin{tabular}{lccc}
\hline
Frequency & Flux density & HPBW  & Reference \\ 
 (MHz)    & (mJy)        & (\arcs) & \\
\hline
73.8  & 162$\pm$100 & 75 & VLSSr \\
147   & 133$\pm$16  & 37.264$\times$23.787 & Our analysis \\
327   & 81$\pm$3    & 54 & WENSS \\
333   & 82$\pm$9.5  & 25 & Our analysis \\
1388  & 46$\pm$3.8  & 3  & Our analysis \\
1400  & 56$\pm$0.45 & 45 & NVSS \\
4850  & 38$\pm$6    & 210 & Becker et al. \\
8400  & 26          & $\sim$0.25 & March{\~a} et al.\\
\hline
\end{tabular}
\end{center}
\end{table}

Table~\ref{tab:radflux} lists our radio flux density measurements for NGC~6338, and previous estimates from the literature. For our own measurements, we used images made with circular restoring beams. At 333~MHz, our GMRT flux density measurement agrees well with the flux from the Westerbork northern sky survey (WENSS). At $\sim$1.4~GHz, our GMRT measurement disagrees with the NRAO VLA Sky Survey (NVSS) flux density, but is in good agreement with a flux density measured from a Faint Images of the Radio Sky at Twenty centimeters \citep[FIRST,][]{Beckeretal95} survey image, suggesting that the larger beam of the NVSS (45\arcs) may be collecting diffuse flux resolved out of the GMRT and FIRST images. The 4.85~GHz measurement was made with the NRAO Green Bank 91m telescope, and its large (3.5\arcm) beam covers several other nearby sources; we therefore consider the flux density to be overestimated. \citep{Marchaetal01} do not quote an uncertainty on their 8.4~GHz flux density measurement, or a beam size. They made the measurement with the VLA in A configuration and we have therefore quoted a typical beam size for that setup.

\begin{figure}
\includegraphics[width=\columnwidth,bb=20 350 570 750]{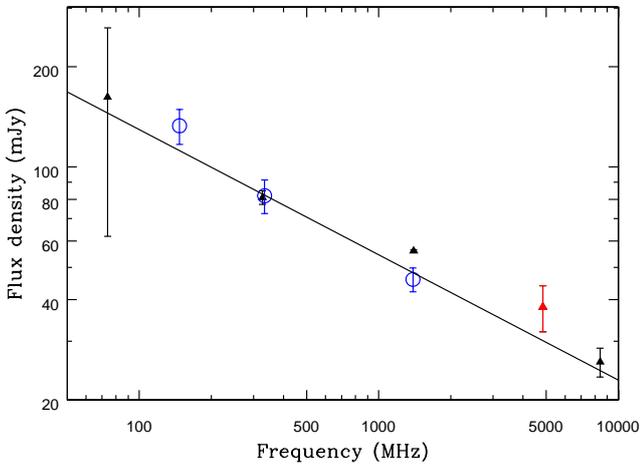}
\caption{\label{fig:spix}Radio spectrum of NGC~6338, with best fitting powerlaw marked by a solid line. Literature flux measurements are marked by solid triangles, our own measurements by open blue circles. The 4.85~GHz single-dish measurement, which must be overestimated since its beam will have included nearby sources, is marked in red.}
\end{figure}

 To determine the spectral index of the source, we fitted a powerlaw to the flux densities, using our 1.39~GHz measurement rather than the FIRST or NVSS fluxes, excluding the 4.85~GHz measurement, and assuming a 10\% uncertainty for the 8.4~GHz measurement. The spectrum is shown in Figure~\ref{fig:spix}. We find a spectral index of $\alpha_{74}^{8400}$=0.38$\pm$0.03\footnote{We assume that flux density $S$ is related to frequency $\nu$ as $S\propto\nu^{-\alpha}$}. This is in reasonable agreement with the spectra indices estimated by \citet{Marchaetal01}, $\alpha_{1400}^{4800}$=0.40 and $\alpha_{1400}^{8400}$=0.44, and that of \citet{Hoganetal15b}, $\alpha_{300MHz}^{150GHz}$=0.41$\pm$0.04, though the latter notes that there may be evidence of variability at high frequencies. Excluding the 8.4~GHz measurement steepens the spectral index slightly, to $\alpha_{74}^{1400}$=0.43$\pm$0.06. 

In addition to the source associated with NGC~6338, two other bright sources are visible in the 1.39~GHz map. A point-like source is visible at the position of 2MASS~J17153152+5722504, a $z$=0.065 galaxy in the background of the group (labelled ``A'' in Fig.~\ref{fig:radio}). A weak X-ray point source is also seen coincident with the radio and optical centroids. To the northeast of this galaxy we see a probable double-lobed source (labelled ``B'') whose position corresponds to that of NVSS~J171538+572322. \citet{Wangetal19} note the presence of an X-ray point source whose position overlaps the brightest part of this source, and associate it with the galaxy SSTSL2~J171538.78+572325.8; we concur with this identification. The NVSS beam is large enough that sources A and B are not resolved separately in the NVSS image. Their combined flux in our GMRT image is consistent with the NVSS flux, confirming that no other source is present; there is no indication of the radio relic suggested by \citet{Wangetal19} at this position.

\subsection{Background group}
\label{sec:BGgrp}
As well as numerous point sources in the field of view, we also identified one extended source, located at 17$^h$14$^m$17\fs 0 +57\degr 31\arcmin 01\farcs9 (beyond the edge of the figures in this paper). This is close to the position of 2MASX~J17141694+5731026, a galaxy at redshift $z$=0.11336. Only a few hundred counts are detected in the ACIS-I observation, but we are able to extract and fit spectra from the long \xmms\ observation. We use a 30\arcs\ radius region, and a local background extracted from a partial annulus to either side of the source, at the same radius from the \xmms\ optical axis.

An absorbed APEC model provides a good fit (reduced $\chi^2$=1.01 for 92 degrees of freedom), with kT=1.06$^{+0.01}_{-0.02}$~keV, 0.85$^{+0.16}_{-0.14}$\Zsol\ abundance, and a fitted redshift $z$=0.108$^{+0.007}_{-0.004}$, in good agreement with the redshift of 2MASX~J17141694+5731026. Adopting the luminosity distance of that object (D$_L$=505~Mpc) the 0.5-7~keV X-ray luminosity is $\sim$2.7$\times$10$^{42}$\ergps\ within $\sim$60~kpc (for a scale of 1.977 kpc/arcsec). This is roughly consistent with expectations from the \LT\ relation. We therefore consider that the source is probably an X-ray bright group, with 2MASX~J17141694+5731026 the dominant galaxy.

\section{Discussion}
\label{sec:disc}

\subsection{Dynamical state of the group}
\label{sec:merger}
From the X-ray imaging and spectral maps, it is clear that the NGC~6338 group is a merging system and, for a group, an unusually violent one. The sharp surface brightness drops on the inward-facing edges of the two cores, and their tails, indicate that both are in motion relative to the surrounding IGM, with sufficient velocity for gas to be stripped from them. The high temperatures and pressures between and around them are most easily explained as shock-heated gas. The low abundance of this gas (0.2-0.4\Zsol) suggests that it originated in the outer parts of the progenitor groups, where only limited enrichment could take place. 

Given that both cores are in motion relative to the surrounding IGM, it seems likely that the merger is between two progenitors of similar mass. The fact that the northern core is smaller and that VII~Zw~700 is clearly a less massive galaxy system than NGC~6338 suggests that the northern progenitor had a lower mass, but the two are clearly comparable. In general terms we can imagine a scenario in which the two groups have fallen together, and are driving shocks into each others IGM, with the motion of the dense cool cores through the shocked gas causing stripping, leaving tails of cooler, enriched gas behind them.

The projected mean temperature of the gas in the region directly between the two cores is 3.60$^{+0.15}_{-0.14}$~keV (a 65\arcs-radius region, excluding the cores, fitting to \chandra\ data). The sound speed in gas of this temperature is $\sim$985\kmps. Since we do not observe shocks ahead of either core, only cold fronts, this suggests a maximum velocity in the plane of the sky, relative to the IGM. However, much of the motion may be along the line of sight. We therefore examine the velocity distribution of the galaxy population.

\subsubsection{Galaxy population distribution}
\begin{figure}
\includegraphics[width=\columnwidth,trim=0 40 0 0]{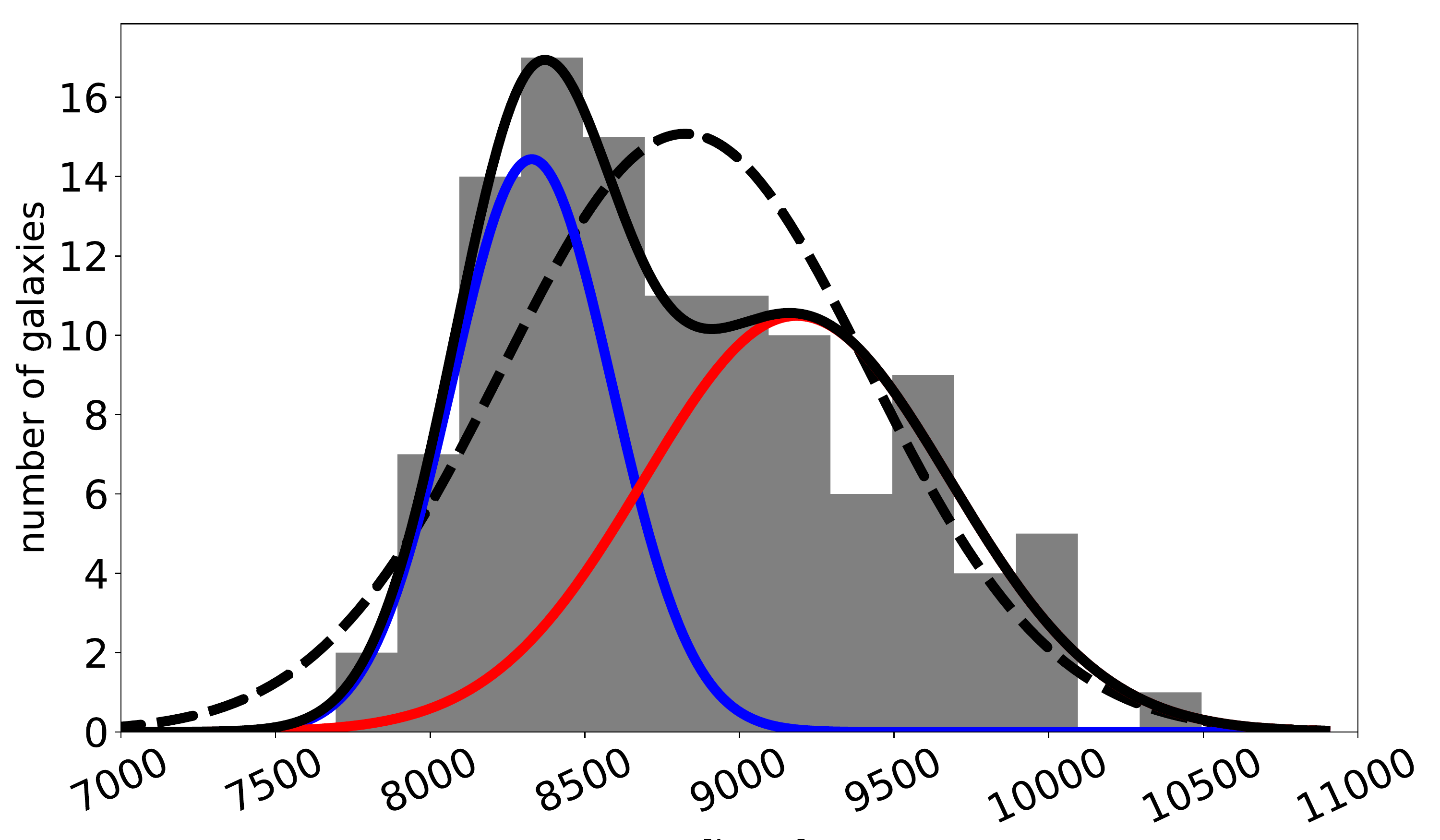}
\caption{\label{fig:vhist}Histogram of galaxy velocities (in \kmps) within 1.5~Mpc and 2000\kmps\ of NGC~6338. The dashed line indicates a single-Gaussian fit to the the data, while the solid lines show the results of the Gaussian mixture modelling, with the red and blue lines indicating the two components and the black line their sum.}
\end{figure}

We identified all galaxies in the NED and HyperLEDA catalogues within a projected distance of 1.5~Mpc (47\farcm3) of NGC~6338 and in the range 7000-11000\kmps. After cross-matching to remove duplicates, a total of 112 galaxies are found within this volume. Figure~\ref{fig:vhist} shows a histogram of the galaxy recession velocities which is clearly skewed, with a peak at $\sim$8400\kmps and a tail extending out to $\sim$10500\kmps.

\begin{figure}
\includegraphics[width=\columnwidth,bb=36 133 577 659]{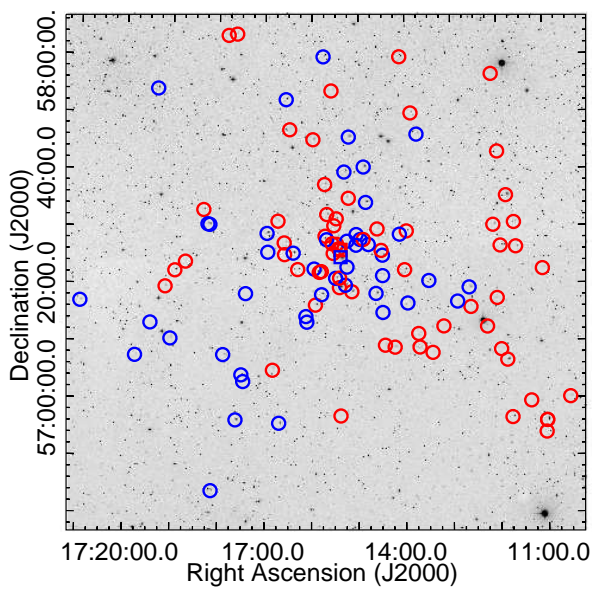}
\vspace{-8mm}
\caption{\label{fig:galdist}Distribution of galaxies within 1.5~Mpc and 2000\kmps\ of the NGC~6338 group. Blue and red circles (squares for the dominant galaxies) indicate members of the low and high velocity subsets, respectively. NGC~6338 and VII~Zw~700 lie in the centre of the image.}
\end{figure}

A single Gaussian is clearly insufficient to model the velocity distribution. 
We therefore apply a Gaussian Mixture Model \citep[GMM, as implemented in the \textsc{scikit-learn} \textsc{python} package][]{Pedregosaetal11} to subdivide the galaxies based on the combination of their positions and velocities. A two component model provides a good approximation of the velocity distribution (see Fig.~\ref{fig:vhist}), and a likelihood ratio test confirms that it is a $>$3$\sigma$ significant improvement on the the single Gaussian. In velocity, the model divides the galaxies into a relatively narrow 47-galaxy low-velocity subset centred at 8328$\pm$38\kmps\ with $\sigma$=261$\pm$31\kmps, and a broader 65-galaxy high velocity component centered at $\sim$9186$\pm$64\kmps with $\sigma$=496$\pm$48\kmps. NGC~6338 (at 8185\kmps) is associated with the low velocity component, VII~Zw~700 (at 9593 and 9630\kmps) with the higher.

Figure~\ref{fig:galdist} shows the spatial distribution of galaxies in the
two GMM components. The two subsets are relatively evenly spread,
suggesting a merger aligned prmarilly along the line of sight. However,
investigation shows that the galaxies with the greatest recession
velocities tend to be located north and west of the system center. This is
consistent with the results of \citet{Wangetal19} who, using only
velocities for a sample of $\sim$80 galaxies within 1~Mpc, found a smaller
high velocity component located north of the center.  We experimented with
adding a third component to the model, but found that its mean velocity and
velocity width were poorly constrained, and that it primarily consisted of
galaxies at one edge of the field. A larger galaxy sample is probably
needed if the galaxy distribution is to be examined in greater depth.

Based on the two-component fit, we have an average line of sight velocity difference between the two components of $\sim$850\kmps, and a velocity difference between the two group-dominant galaxies of $\sim$1400\kmps. These are very large velocities for groups, and given the gas tails observed in the X-ray, they can only be lower limits on the true velocity of the merger.

\subsubsection{Comparison with simulations}
\label{sec:sims}

To better understand the dynamics of the merger, we compared our observations with simulations in the Galaxy Cluster Merger Catalog\footnote{http://gcmc.hub.yt} \citep{ZuHoneetal18}. Among other products, this provides projected temperature distributions and simulated X-ray images for a parameter space exploration of cluster mergers, with data for mergers of different mass ratio, impact parameter, and angle to the line of sight \citep[see][for a detailed description of the simulations]{ZuHone11}. The simulations assume a primary cluster mass of 6$\times$10$^{14}$\Msol, much more massive than our system, but while this means the absolute temperatures are not comparable, the temperature and surface brightness structures in the simulations can still provide a guide to the state of NGC~6338. 

\begin{figure}
\includegraphics[width=\columnwidth,bb=30 210 495 680]{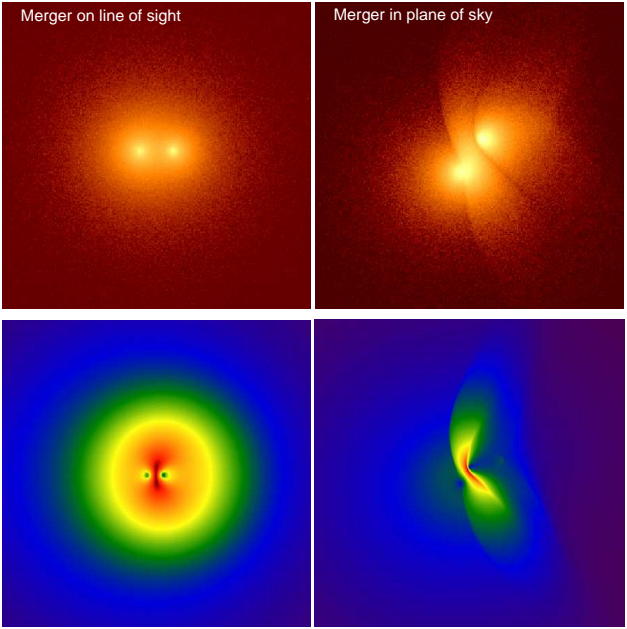}
\caption{\label{fig:sim}Simulated X-ray images (\textit{top}) and projected temperature distributions (\textit{bottom}) from the Galaxy Cluster Merger Catalog, showing a 3:1 mass ratio merger with moderate impact parameter, projected along the line of sight (\textit{left}) or in the plane of the sky (\textit{right}). The temperature scale runs from blue (cold) to red (hot); in the \textit{lower left} panel the temperature range is roughly a factor of 4, and the dark linear feature between the two cores is dark red, indicating the highest temperatures. Note that the angular and temperature scales are arbitrary; the images are only intended to show examples of morphology. In all panels the less massive group is on the right; in the right hand panels it is falling in from the right hand side. Note the presence of sharp surface brightness edges associated with the shock front in the right-hand panels, and their absence for the ine of sight merger.}
\end{figure}

An initial examination confirms our basic understanding of the merger. It is likely to have a near-equal mass ratio, and is probably close to core passage (i.e., the closest approach of the two cores) and certainly not far past it. Figure~\ref{fig:sim} shows example X-ray surface brightness and temperature distributions for a 3:1 mass ratio merger with a moderate impact parameter (500~kpc). A merger in the plane of the sky produces prominent shock fronts with sharp surface brightness edges ahead of each core and extending to large distances perpendicular to the direction of motion. In the case of an equal-mass merger with low impact parameter, these are so prominent that the X-ray surface brightness distribution of the cluster becomes cross-shaped. We do not see any such features in either the \chandra\ (see Figure~\ref{fig:introims}) or \xmms\ images. 

Mergers along the line of sight, with some impact parameter separating the cores, produce no surface brightness edges; the shock fronts are travelling along the line of sight and we are looking through them. Instead the simulations predict a smoothly elliptical X-ray surface brightness distribution, with a surface brightness dip between the two cores. The projected temperature is enhanced between the two cores and in broad regions to either side of the axis between them, forming a figure-eight shape of high temperatures. The shapes of these features depends on the parameters of the merger. For equal mass mergers, even at large impact parameters, the axis of the high temperature region is perpendicular to the merger axis. For more unequal mergers, the high temperatures bend back toward the direction of infall of the smaller progenitor, and the surface brightness gap between the cores may also be V-shaped. High temperature shocked gas will extend throughout the central part of the merger, but along lines of sight which pass through the cores and tails, the projected temperature will appear lower. The highest temperatures will be seen along lines of sight that pass outside the cool cores.

The angle of the merger axis to the line of sight, the impact parameter and the separation of the two cores can all trade off against one another to produce similar projected temperature and surface brightness distributions. However, the simulations support a scenario in which the merger is occurring along an axis close to, but (since we see stripped tails) not perfectly aligned with, the line of sight. The mass ratio is likely 3:1 or closer.

\subsubsection{Merger velocity and energy injection}
Since the shock fronts are likely travelling in the line of sight, we cannot directly measure the velocity of the merger, but we can make an estimate based on the temperatures. The shape of the temperature profile of NGC~6338 is typical of cool core groups, with a temperature peak at moderate radii and a decline in the outskirts. Beyond 0.3$\times$R$_{500}$ such profiles typically only have mild negative gradients. If we take the scaled temperature profiles of a sample of cool core groups \citep[e.g,][]{OSullivanetal17}, the temperature peak is usually $\leq$1.5$\times$ the temperature outside 0.3$\times$R$_{500}$. At large radii in NGC~6338 we find temperatures $\sim$1.3~keV. That suggests that the progenitors are unlikely to have contained gas hotter than $\sim$2~keV before the merger. We now observe peak temperatures of 5~keV in the temperature maps. For a pre-shock temperature of 2~keV (or 1.3~keV) this implies a temperature increase of a factor 2.5 (or 3.85), requiring a shock of Mach number $\mathcal{M}$=2.3 (or $\mathcal{M}$=3.1). This would imply a merger velocity of $\sim$1700-1800\kmps, comparable to but exceeding the line-of-sight velocity difference between the two BGGs. We must therefore conclude that the direction of the merger is close to the line of sight, that it has driven quite powerful shocks into the IGM of the two progenitor groups, and that the two cores are still moving at nearly the sound speed of this shocked gas.

Estimating the energy injected into the IGM by the merger is hampered by the fact that we do not know the true extent of the shock heated region. We can estimate a lower limit by making two assumptions: a) that the pre-shock temperature was either 1.3 or 2~keV, and b) that the shocked volume can be approximated as a sphere whose radius is set to the point on an east-west temperature profile (excluding both cores and tails) where the temperature falls to that pre-shock temperature ($\sim$350\arcs/185~kpc for 2~keV). These assumptions are conservative, since the shock will likely extend to larger radii, and the volume may be much deeper along the line of sight, but will give an approximation of the energy injection into the core. Depending on the pre-shock temperature chosen, we find the likely energy injection within this region to be at least 2-8$\times$10$^{57}$~erg.

\subsection{Surface brightness discontinuities}
\label{sec:fronts}

As reported by \citet{Wangetal19}, there are sharp changes in surface brightness and temperature at the southern edge of the north core, and the north edge of the south core. These appear to be cold fronts, suggesting the incursion of the cool cores into the shock heated part of the IGM. Our deeper \chandra\ ACIS-S observation reveals these edges to be structured, with the sharpest discontinuities only found in relatively narrow angular ranges. Our pseudo-pressure map (Figure~\ref{fig:PS}) shows no appreciable change in pressure across the cold fronts, as expected, while the pseudo-entropy map shows significantly higher entropies outside the fronts.

The properties of the gas on either side of a cold front can be used to estimate the velocity of the intruding cold material, as described by \citet{MarkevitchVikhlinin07}. \citet{Wangetal19} estimate the pressure change across the fronts and use this to place limits on the plane-of-sky velocities of the cores, relative to the surrounding IGM. However, since a large fraction of the velocity of the merger is along the line of sight, the cold fronts we observe are likely to be at the sides of the cores. It is therefore unclear whether the assumptions of the analysis usually applied to cold fronts (e.g., free streaming gas approaching a dense body with a stagnation zone ahead of its leading edge) are valid in these circumstances. We have carried out the analysis, as a check on our interpretation of the pseudo-pressure map and to determine whether our deeper \chandra\ observations can reveal any pressure offset, but we note that the estimates of the plane-of-sky velocity should only be considered as indicative, not as definitive measurements.

\begin{figure}
\includegraphics[width=\columnwidth,trim=0 0 0 41,clip]{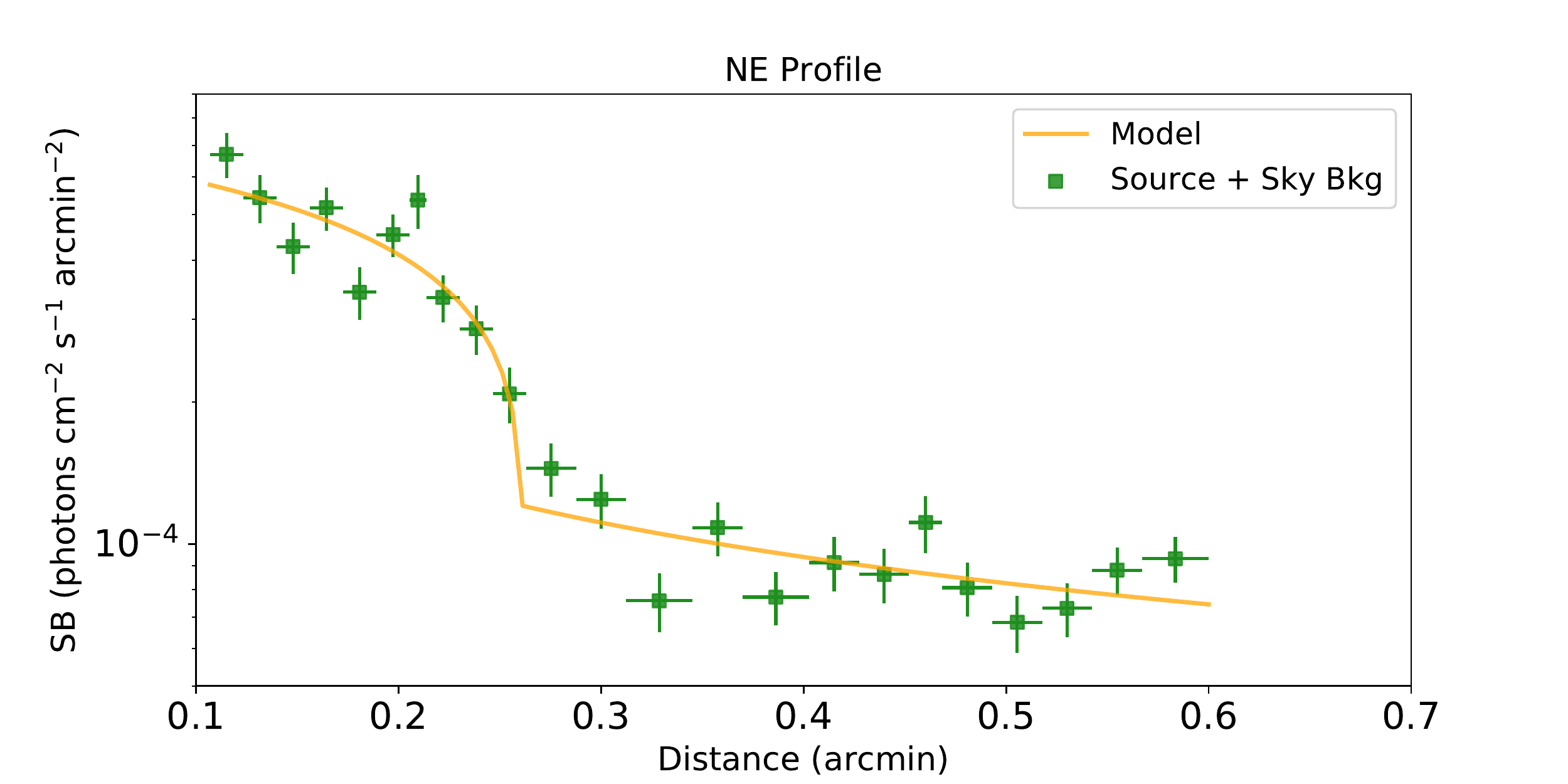}
\includegraphics[width=\columnwidth,trim=0 0 0 41,clip]{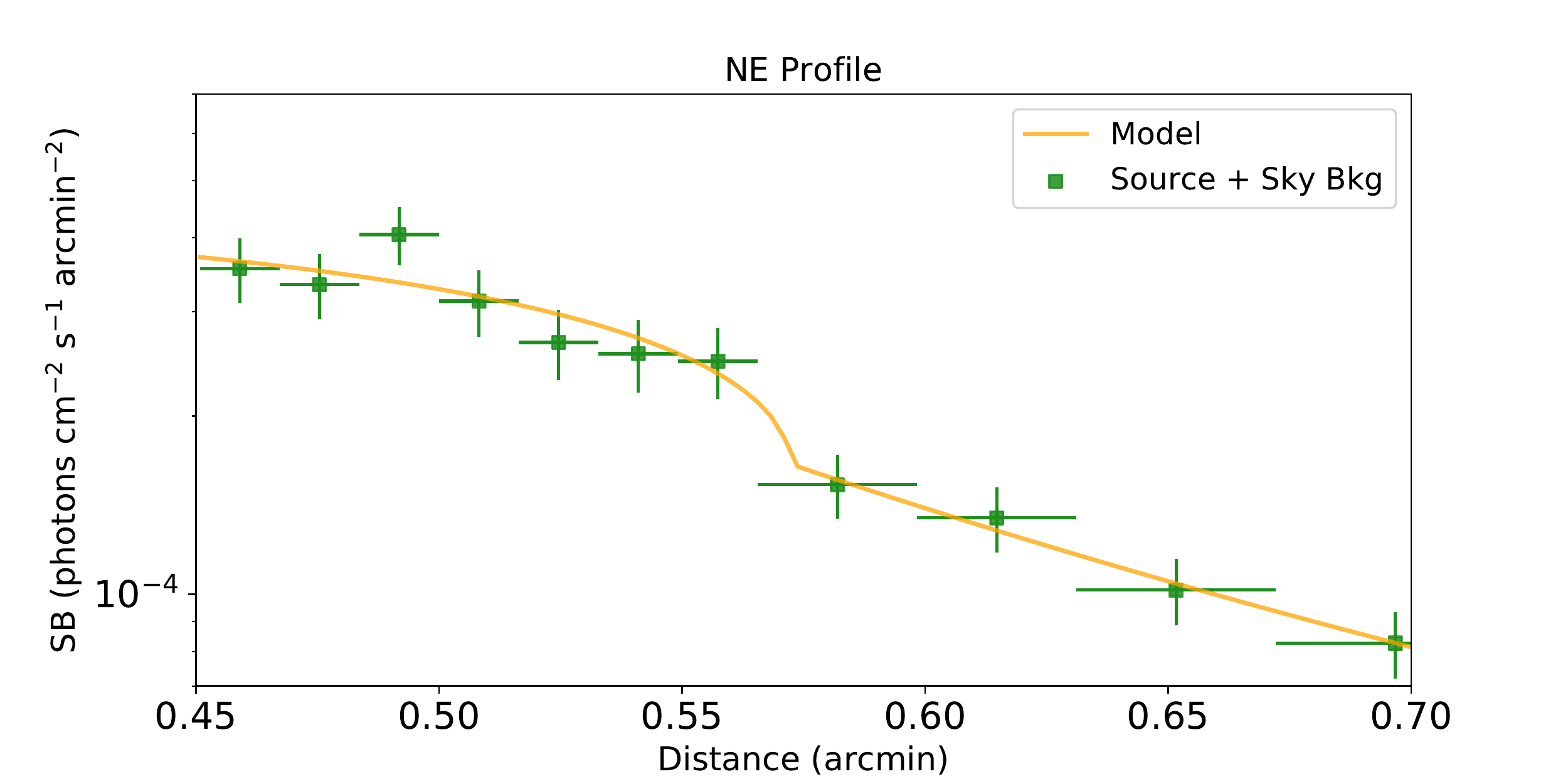}
\caption{\label{fig:fronts}Surface brightness profiles extracted across the north (\textit{upper panel}) and south (\textit{lower panel}) cold fronts, fitted with two-component powerlaw density models.}
\end{figure}

We extracted surface brightness profiles from relatively narrow sectors (opening angle 53\degree\ for the northern front, 20\degree\ for the southern) extending across the sections of each front where the sharpest boundary was observed. We model the surface brightness profiles using projected density models, consisting of two powerlaw components with a jump at the front boundary, as in e.g., \citet{Ogreanetal16}. The models were integrated assuming a spherical geometry. Fitting was performed using the \textsc{PyXel}\footnote{https://github.com/gogrean/PyXel} code \citep{Ogrean17}. Figure~\ref{fig:fronts} shows the radial profiles and model fits. The discontinuities are clear in both profiles, and the fitted models have density jumps of factor 3.63$^{+0.65}_{-0.54}$ across the northern front and 1.39$\pm$0.25 across the southern front.

To measure the temperature inside each front we used partial annuli 8-10\arcs\ ($\sim$4-5~kpc) wide in the same narrow sectors, using a local background region to subtract off IGM emission along the line of sight. Each contains $\sim$600 net counts. We find kT=2.81$^{+0.38}_{-0.41}$~keV inside the south front, and kT=1.83$^{+0.21}_{-0.15}$~keV inside the northern front, in good agreement with the temperature maps. Estimating the true temperature of the gas through which the cores are moving is difficult given the geometry of the system. We chose to use the temperature from the local background region, which was made up of partial annuli to the east and west of the cores, $\sim$100\arcs\ from the midpoint between the two, in the hot shocked region of the temperature map. The temperature of these regions is 3.95$\pm$0.25~keV. 

Combining the measured temperature and density jumps, we find that the pressure ratio across the southern front is P$_{\rm in}$/P$_{\rm out}$=0.99$\pm$0.23. If we treat the discontinuities as simple cold fronts, and take the gas outside the front to be freely streaming subsonically past the the core, the velocity of the gas is related to the pressure ratio by \citep{LandauLifshitz59}:

\begin{equation}
\frac{\rm P_{in}}{\rm P_{out}} = \left(1+\frac{\gamma-1}{2}\mathcal{M}^2\right)^\frac{\gamma}{\gamma-1},
\end{equation}

\noindent where $\gamma$=5/3 is the adiabatic index of the gas and $\mathcal{M}$ is the Mach number. For the south core, the 1$\sigma$ upper limit on the pressure ratio is equivalent to a limit on the Mach number of $\mathcal{M}$$\leq$0.5, implying an angle of motion $\leq$27\degree\ to the line of sight. For the northern core the pressure ratio is considerably higher, P$_{\rm in}$/P$_{\rm out}$=1.67$\pm$0.35, mainly because of the large density jump. This pressure ratio is equivalent to $\mathcal{M}$=0.83, but the uncertainties mean that this value is consistent with ratios of 1 (no motion in the plane of the sky) and 2 (plane of sky velocity equal to the sound speed) within 2$\sigma$. We are therefore unable to place any strong constraint on the motion of the north core. 

\citet{Wangetal19} find $\mathcal{M}$$\leq$0.6 (at 90\% confidence) for the south core, in good agreement with our value, but $\mathcal{M}$$\leq$0.8 for the north core, consistent with but lower than our estimate. Our high pressure ratio is largely driven by the large density jump across the front. The density profile crosses the bright rim around the southern cavity, which may be material compressed or uplifted by the expansion of the cavity. We also note that our surface brightness profiles are tailored to match the regions of sharpest discontinuity, so cover a much narrower part of each front than those used by Wang et al. We also use smaller spectral extraction regions, to measure the temperature immediately inside the fronts without contamination from the cooler gas in the cores. Wang et al. also use regions at the trailing edge of each core (i.e., at the base of the tails) as a proxy for the temperature of the free streaming gas outside the fronts. The temperatures of these regions may be comparable to that of the pre-shocked gas ($\sim$2~keV), but it is unclear that they are applicable to the gas flowing past the cores.


\subsection{Ram-pressure stripping}
The X-ray images and spectral maps of the group show tails of cooler, lower entropy, metal enriched gas behind both cores. Along with the offset between the gas and dominant galaxy in the north core, the tails are evidence that the motion of the cores through the surrounding medium is rapid enough to lead to extensive stripping. The lower mass of the northern progenitor explains the somewhat greater extent of the northern tail, as well as the X-ray/optical offset; the more massive southern core has thus far remained undisrupted. While the southern tail appears to have a relatively simple structure, Figure~\ref{fig:tails} shows what may be evidence of hydrodynamic instabilities in the north tail. The eastern edge of the tail is strongly curved, and there is a large ($\sim$19~kpc across, in projection) depression in the surface brightness of the western edge. Although this region is crossed by an ACIS-I chip gap and bad columns in some of the ACIS-S3 data, these structures are not the cause of the surface brightness deficit, which is larger than them and has rounded rather than linear boundaries. Our entropy map (Figure~\ref{fig:PS}) shows a spur of higher entropies extending into the tail from the surrounding shocked gas at the position of the depression. Dividing the tail into east and west halves and extracting spectra along its length, we find that this entropy feature corresponds to a $\sim$2$\sigma$ significant jump in temperature. This suggests that the surface brightness depression is caused by an intrusion of hotter, less dense gas from the surrounding IGM into the stripped tail.

The depression and curved structure of the tail resemble features seen in other clusters and groups \citep{Mazzottaetal02,Russelletal12b,Russelletal14} which are believed to arise from turbulent mixing. The growth of such structures can be suppressed by viscosity and by magnetic fields. Following \citet{Suetal17}, we use the measured gas properties to place limits on these parameters.

For a shear flow of Mach number $\mathcal{M}$, \citet{Roedigeretal13} show that for Kelvin-Helmholtz instabilities to form, viscosity must be less than some fraction of the Spitzer value, $f_\nu$, where:

\begin{multline}
f_\nu < \frac{10}{16\sqrt{\Delta}}\cdot\frac{\mathcal{M}v_{sound}}{400\kmps}\cdot\frac{l}{10~\mathrm{kpc}}\cdot \\ \frac{n_{e,h}}{10^{-3}\pcmcu}\cdot\Big( \frac{kT_h}{2.4~\mathrm{keV}}\Big)^{-5/2},
\label{eqn:visc}
\end{multline}

\noindent $v_{sound}$ is the sound speed in the flow, $kT_h$ and $n_{e,h}$ are the temperature and electron density in the flow, $l$ is the length scale of the instability, and $\Delta$=($\rho_h$+$\rho_c$)$^2$/$\rho_h\rho_c$ describes the contrast in gas density across the interface. Adopting $kT_h$=3.6~keV, $n_{e,h}$=2.5$\times$10$^{-3}$\pcmcu and $\mathcal{M}$=0.71, we find $f_\nu$$<$0.43. This is compatible with recent results showing sub-Spitzer viscosities in galaxy clusters \citep[e.g.,][]{Suetal17,Ichinoheetal17}.

Similarly, we can place a limit on the strength of an ordered magnetic field given that it is insufficient to suppress the formation of instability structures,

\begin{equation}
\frac{B^2}{8\pi} < \frac{\gamma\mathcal{M}^2}{1+T_c/T_h}.P,
\label{eqn:magstable}
\end{equation}

\noindent where $B$ is the magnetic field strength, $T_c$ and $T_h$ are the temperatures in the tail and flow respectively, and $P$ is the thermal pressure in the flow \citep{Vikhlininetal01b}. From this we can place a limit of $B$$<$19.1~$\mu$G, which is compatible with estimates $<$10~$\mu$G in other systems \citep[e.g.,][]{Suetal17}.

Our limits on viscosity and magnetic field strength are dependent on the angle to the line of sight of the merger. We have conservatively assumed that the projected distance of the structure from the north core is similar to its true distance, but as the merger is probably aligned along the line of sight, the true distance may be greater. A more distant structure would be surrounded by less dense gas, at a lower pressure, making the limits given above more restrictive (eqns.~\ref{eqn:visc} and \ref{eqn:magstable}). An accurate determination of the merger orientation would therefore place tighter constraints on viscosity and magnetic field strength in the system.




\subsection{Cooling and AGN feedback}

\subsubsection{Cavities and jet powers}
\label{sec:cav}
Using the ACIS-I observation, \citet{Pandgeetal12} identified cavities northeast and southwest of the nucleus of NGC~6338 and estimated that they were inflated by jets of power $\sim$6-10$\times$10$^{42}$\ergps over timescales of $\sim$60-120~Myr. They reported that the NVSS contours of the central radio source overlap the cavities. However, those contours are consistent with emission from a point source at the galaxy centroid convolved with the large NVSS beam, so this does not indicate the presence of radio emission from the cavities. 

Figures~\ref{fig:Score} and \ref{fig:Ncore} show \chandra\ images and residual maps for the two cores. In the residual map of the south core, we see the same surface brightness deficits as Pandge et al., though deeper imaging means these are better defined. In particular, the "rim" structure helps constrain the size of the northeastern deficit. We assume the cavities to be ellipsoidal and estimated their size based on the residual map and unsharp mask images. The northwest ellipse overlaps the AGN. This may indicate that the cavity is not in fact ellipsoidal, or may be an orientation effect.

Beyond the ends of the east and south branches of the filamentary structure, we see a depression in the X-ray surface brightness enclosed by a brighter rim; this may be another cavity. There are also areas of lower X-ray surface brightness around the outer part of the West branch of the filaments. This could potentially indicate another cavity, but its size and shape are less clear, so we consider only the southeast structure in our calculations.  


Figure~\ref{fig:Ncore} shows the residual map of the north core. Here we also see some potential cavities, most obviously the surface brightness deficit on the southwest side between the bar and rim. To the north of the bar the image is more difficult to interpret, with a large surface brightness deficit northeast of the bar, but extending around the west knot. We interpret the region of greatest surface brightness deficit as a potential cavity, with the remaining structure likely a product of gas stripping from the core to form the tail. The two potential cavities have different sizes and orientations, and like all the structures in the north core, are offset from the optical centre of MCG+10-24-117. If these are indeed cavities, it seems likely that they formed when the centre of the X-ray bar was collocated with the nucleus of MCG+10-24-117, and for our calculations we assume the centre of the bar as the origin of the cavities.

For each candidate cavity we estimate the enthalpy, defined as 4$PV$ where $P$ is the deprojected IGM pressure at the midpoint of the cavity, and $V$ is its volume. Cavities are assumed to be ellipsoidal, with an axis of symmetry along the (approximate) line of the jets which would have formed them. Uncertainties on the enthalpy are estimated following the approach described in \citet{OSullivanetal11b}. We estimate the cavity ages based on their buoyant rise time, sonic expansion timescale and refill time \citep[see e.g.,][]{Birzanetal04}, and determine the mechanical jet powers necessary to inflate the cavities by dividing their enthalpy by these timescales. The results of these calculations are shown in Table~\ref{tab:cav}. The lack of complete rims of compressed gas, or of shocked material associated with the cavities, suggests that their motions are subsonic.

\begin{table*}
\caption{\label{tab:cav} Cavity dimensions, enthalpies and timescales, and the resulting jet mechanical power estimates. The cavity semi-major axes are given by r$_{maj}$ and r$_{min}$, and their mean distance from the origin by R.} 
\begin{center}
\begin{tabular}{lcccccccccccc}
\hline
Core & cavity & r$_{maj}$ & r$_{min}$ & R & Pressure & 4$PV$ & \multicolumn{3}{c}{cavity age} & \multicolumn{3}{c}{Jet power} \\
 & & & & & & & t$_{sonic}$ & t$_{buoy}$ & t$_{refill}$ & P$_{sonic}$ & P$_{buoy}$ & P$_{refill}$ \\
     &        & (arcs)   & (arcs)   & (arcs) & (10$^{-11}$~erg~cm$^{-3}$) & (10$^{56}$~erg) & \multicolumn{3}{c}{(Myr)} & \multicolumn{3}{c}{(10$^{42}$\ergps)}\\
\hline \\[-3mm]
S & inner NE & 8.7  & 6.1  & 7.5  & 24.3$\pm$1.3 & 81.5$^{+38.6}_{-19.3}$ & 7.3 & 5.6 & 19.3 & 35.4$^{+16.7}_{-8.4}$ & 46.3$^{+21.9}_{-11.0}$ & 13.4$^{+6.4}_{-3.2}$ \\[+1mm]
  & inner SW & 8.0  & 6.1  & 12   & 14.4$\pm$0.7 & 40.8$^{+15.4}_{-10.1}$ & 11.4 & 14.7 & 30.8 & 11.3$_{-2.8}^{+4.3}$ & 8.80$^{+3.32}_{-2.17}$ & 4.20$^{+1.58}_{-1.04}$\\[+1mm]
  & outer SE & 12.3 & 7.6  & 34.5 & 6.21$^{+0.26}_{-0.27}$ & 51.9$^{+33.5}_{-11.5}$ & 25.8 & 33.5 & 32.7 & 6.38$^{+4.11}_{-1.41}$ & 4.91$^{+3.17}_{-1.09}$ & 5.03$^{+3.25}_{-1.11}$ \\[+1mm]
N & NE & 7.5 & 4.6  & 10.5 & 0.79$^{+0.21}_{-0.24}$ & 0.96$^{+0.67}_{-0.43}$ & 44.0 & 41.7 & 95.5 & 0.07$^{+0.05}_{-0.03}$ & 0.07$^{+0.05}_{-0.03}$ & 0.03$^{+0.02}_{-0.01}$ \\[+1mm]
  & SW & 8.9 & 3.6  & 6.5  & 0.79$^{+0.21}_{-0.24}$ & 1.7$^{+2.6}_{-0.76}$ & 27.3 & 29.3 & 82.8 & 0.20$^{+0.30}_{-0.09}$ & 0.19$^{+0.28}_{-0.08}$ & 0.07$^{+0.10}_{-0.03}$ \\[+1mm]
\hline
\end{tabular}
\end{center}
\end{table*}

Within the uncertainties imposed by the timescale, our jet power estimate for the inner SW cavity in the south core agrees reasonably well with that of \citet{Pandgeetal12}. Our estimate for the inner NE cavity is larger, partly because we adopt a larger cavity size and partly because its small distance from the centroid places it in a higher pressure bin. For the outer cavity, timescales are longer and jet powers consequently smaller, but these may be underestimated if the cavity has been confined by external pressure associated with the south core's motion through the IGM. If an equivalent outer NW cavity is present, the lack of a clear outer X-ray rim may indicate that it was able to expand further. In general the cavity enthalpies and jet powers are fairly typical for a group or poor cluster, and the jet powers exceed the bolometric X-ray luminosity in the core ($\sim$10$^{42}$\ergps\ within 41\arcs/21.9~kpc) by about an order of magnitude.

In the north core, the enthalpies and jet powers are significantly smaller, reflecting the lower gas pressures. However, the jet powers of a few $\times$10$^{40}$-10$^{41}$ are a little more closely matched to the X-ray luminosity of the core ($\sim$3$\times$10$^{40}$\ergps\ within 22\arcs/11.7~kpc). This is perhaps unsurprising given that the northern core contains less gas, is associated with a smaller dominant galaxy, and seems to have been more strongly affected by interaction with the IGM. 

Identifying cavities from X-ray images is difficult and subjective, particularly in complex systems such as this one. In the north core, the SW cavity seems the more reliable, but the size and shape of the NE cavity are unavoidably uncertain. In the south core, the lack of radio emission associated with any of the cavities is puzzling, particularly for the inner cavities whose dynamical ages are only $\sim$5-30~Myr. If both sets of cavities are real, this implies that the AGN jet axis has changed in the last $\sim$15~Myr. The outer southern cavity and filaments lie close to the minor axis of NGC~6338, while the axis of the inner cavities is similar to, but offset from, the galaxy major axis. \citet{Gomesetal16b} show that the galaxy is rotating about its major axis, so the presence of outer cavities would imply a past offset between the stellar rotation axis and that of the central supermassive black hole (SMBH). On the other hand, our GMRT 1.39~GHz data show a hint of extension to the southwest, suggesting that any residual jet emission may be aligned with the outer rather than the inner cavities. Deeper radio observations may offer the best chance to resolve the question of cavity scales and jet orientation in the south core.

\subsubsection{Cooling triggers}
As described in Section~\ref{sec:intro}, previous studies of the cooling regions of galaxies, groups and clusters have shown strong correlations between jet activity, the presence of star formation, molecular gas and H$\alpha$ nebulae, and rapid cooling in the X-ray emitting gas. In particular, it has been shown that nebular emission and star formation typically occur in systems with t$_{cool}$$<$1~Gyr at 10~kpc and/or a minimum value of the ratio t$_{cool}$/t$_{ff}$ that is $<$35, though there are now strong arguments that the latter criterion is driven by the former \citep{Hoganetal17}. The extent of H$\alpha$ nebulae is also known to correlate with the radius at which t$_{cool}$/t$_{ff}$ reaches its minimum \citep{Hoganetal17}. 

The H$\alpha$ nebula in NGC~6338 is relatively luminous compared with other nearby giant elliptical galaxies \citep{Lakhchauraetal18}, and is well correlated with the coolest X-ray structures. It meets both the t$_{cool}$ and t$_{cool}$/t$_{ff}$ criteria for nebular emission, having t$_{cool}$=817$\pm$77~Myr at 10~kpc, and minimum t$_{cool}$/t$_{ff}$=19.7$\pm$3.2 \citep[we estimate t$_{ff}$ from the stellar velocity dispersion as in][]{Voitetal15}. The gas entropy at 10~kpc is 20.8$^{+0.06}_{-0.08}$\kevcmsq, also typical of systems with extended nebular emission \citep{Lakhchauraetal18}. The extent of the H$\alpha$ filaments ($\sim$9~kpc) is comparable to the radius at which the cooling time falls below 1~Gyr ($\sim$11~kpc). The minimum value of t$_{cool}$/t$_{ff}$ is found in the 4.1-7.2~kpc bin of our deprojected spectral profile, but the value in the 7.2-10.8~kpc bin is only 26.8$\pm$4.

The radio AGN and its cavities, and the ongoing merger, are potential sources of gas perturbations, if such are necessary to trigger cooling and condensation. It is possible that the inner cavities have played a role in shaping the filamentary structures, by compressing the cool gas between them as they inflated. Alternatively, it is possible that the filaments were formed by the uplift of gas from the core by the buoyant rise of the outer cavities. The maximum mass of gas which could be uplifted by each cavity is equal to the mass of gas it displaces, and simulations suggest \citep{Popeetal10} that in practice the uplifted mass might be only $\sim$50 per cent. of the displaced mass. We calculate the displaced mass for the outer southern cavity to be $\sim$1.9$\times$10$^8$\Msol, compared to gas masses of 1.7$\times$10$^8$\Msol\ and 0.8$\times$10$^8$\Msol\ for the south and east branches of the filaments, respectively. The cavity therefore appears to be too small to have formed the filaments, by a factor of $\sim$2.6. However, we note this calculation assumes a) that the X-ray gas completely fills the cylindrical volume we have assumed, b) that the filaments and cavities are in the plane of the sky, and c) that the cavity is an oblate ellipsoid. If the filling factor of the filament is $<$1, the gas mass to be uplifted will be reduced. Conversely, if the filaments and cavities are at an angle to the line of sight, the cavity's displaced mass will be reduced, and the volume and mass of the filament increased, making uplift more difficult. A cavity that is more elongated along the line of sight will uplift more gas. It is therefore unclear whether the filamentary structures can consist purely of uplifted gas. Uplift of a smaller mass of gas, which then triggered cooling instabilities, might be more plausible.

In VII~Zw~700, the relationship between the coolest gas and the galaxy is unclear. The H$\alpha$ clump is $\sim$5~kpc from the galaxy centroids (in projection). Even if, as we suspect, the X-ray bar has been pushed back by ram-pressure, the H$\alpha$ clump would never have been located at the galaxy center, but off to one side. Our deprojection analysis suggests that the cooling time of the gas is $>$1~Gyr at all radii, though this approach is not ideal given the asymmetric gas structure. The spectral maps show the coolest, densest, lowest entropy gas is located at either end of the bar. One possibility is that, at some point in the past, VII~Zw~700 hosted a structure similar to that in NGC~6338, an X-ray and H$\alpha$ bar formed from cooling material, but that an AGN outburst and/or the impact of the merger disrupted the cooling and heated the bar enough to stop cooling in its center.

\subsection{Mass estimate}
\label{sec:mass}
Given that the NGC~6338 system seems to be a group-group merger, our ability to estimate an accurate mass from the X-ray data is limited. The necessary assumption of hydrostatic equilibrium probably does not hold for the entire IGM, and the underlying dark matter distribution is not relaxed. Nonetheless, we estimate a hydrostatic mass, but note that some degree of inaccuracy is inevitable.

We follow the procedure of \citet{SchellenbergerReiprich17a} to estimate the mass. This involves fitting a mass model to the projected temperature and surface brightness profiles, assuming an NFW form for the mass profile \citep{NavarroFW95} and taking the concentration-mass ($c-M$) relation of \citet{Bhattacharyaetal13} as a prior. We used the same regions used in our deprojection analysis, excluding the north core and tail. We fitted the combined \chandra\ and \xmms\ temperature profiles \citep[correcting for systematic differences between the \xmms\ and \chandra\ temperatures using the cross-calibration conversion of][]{Schellenbergeretal15}, and the \chandra\ surface brightness profile, excluding the central 5\arcs\ of the south core, which is affected by the cavities and filamentary structures. The gas density profile is normalized to match the observed emissivity in an outer annulus. Table~\ref{tab:mass} shows our results.

\begin{table*}
\caption{\label{tab:mass}Hydrostatic estimates of total mass, gas mass and concentration at the fiducial radii R$_{500}$ and R$_{200}$.}
\begin{center}
\begin{tabular}{lcccccccc}
\hline
 & $M_{tot,500}$ & $R_{500}$ & $c_{500}$ & $M_{gas,500}$ & $M_{tot,200}$ & $R_{200}$ & $c_{200}$ & $M_{gas,200}$ \\
 & (10$^{13}$\Msol) & (kpc) & & (10$^{13}$\Msol) & (10$^{13}$\Msol) & (kpc) & & (10$^{13}$\Msol) \\
\hline
Full profile & 5.42$\pm$0.05 & 570$\pm$10 & 14.70$^{+0.65}_{-0.49}$ & - & 6.38$^{+0.05}_{-0.06}$ & 820$\pm$10 & 21.06$^{+0.93}_{-0.71}$ & - \\
Excl. core, outer bin & 9.05$^{+0.14}_{-0.11}$ & 680$^{+0}_{-0}$ & 9.66$^{+0.30}_{-0.36}$ & 1.21$\pm$0.01 & 11.00$^{+0.17}_{-0.13}$ & 980$^{+10}_{-0}$ & 13.99$^{+0.43}_{-0.51}$ & 1.82$\pm$0.01 \\
\hline 
\end{tabular}
\end{center}
\end{table*}

We find very small statistical uncertainties on both total mass and gas mass. These parameters are most strongly constrained by the temperature and emissivity in the outer annuli of the \xmms\ profile. The \xmms\ ESAS fitting produces quite small uncertainties on these values in the outer bins, for several reasons. The background model is well constrained, since we fit to data from the entire \xmms\ field of view, plus a large-radius \rosat\ All-Sky Survey (RASS) spectrum to help define the contribution from soft emission. We selected large outer annuli to ensure high signal-to-noise ratios. Our outermost annulus has radii 370-475~kpc (11.7-15.0\arcm), equivalent to $\sim$0.6$\times$R$_{500}$. The temperature in the outer bin is also cool enough that emission from the $\sim$1~keV Fe L-shell complex is a prominent feature, and likely provides strong constraints on the temperature. However, the outermost bin is not deprojected, and will contain projected emission from gas at larger radii, which may bias its temperature. We therefore performed the analysis again with the outermost bin excluded. This gives a somewhat higher mass estimate, close to the \textit{Planck} SZ estimate (M$_{500}$=1.01$\pm$0.12 $\times$10$^{14}$\Msol), and in reasonable agreement with prior X-ray mass estimates based on the \chandra\ ACIS-I data \citep{Sunetal09} and the 2014 \xmm\ observation \citep{Lovisarietal15}. The difference between our two estimates gives some indication of the systematic uncertainties involved. 

Given the agreement between our larger estimate and the SZ mass, we consider that despite the uncertainties, this result at least gives us a context for the system: NGC~6338 is a merger between two groups which, when it has had time to relax, will form a galaxy cluster.

\subsection{Prospects for future missions}
\label{sec:future}
While our \chandra\ and \xmms\ observations have allowed us to characterize several aspects of the NGC~6338 system, a number of issues remain unaddressed, most notably the angle of the merger axis to the line of sight. A comparison with tailored simulations might help constrain this, but with the current data some degree of uncertainty is inevitable, given the trade-offs between merger angle, impact parameter and merger stage.

The \textit{Athena} X-ray observatory \citep{Nandraetal13}, expected to launch in the mid 2030s, may offer a opportunity to resolve this issue. In addition to a very large collecting area (1.4~m$^2$ at 1~keV), \textit{Athena} will carry an X-ray Integral Field Unit \citep[X-IFU][]{Barretetal18}, a cryogenically cooled imaging spectrograph with a $\sim$5\arcm-diameter field of view, 5\arcs\ pixels, and 2.5~eV spectral resolution. This unprecedented combination of spatial and spectral resolution offers the opportunity to map the velocity structure of the IGM by measuring the redshift of X-ray emission lines. \citet{Roncarellietal18} have demonstrated this potential using simulations of a massive cluster at $z$=0.1, showing that deep X-IFU observations will accurately map both the velocity structure of the ICM and the line broadening caused by turbulence. Measuring the velocity structure of the NGC~6338 system would provide a powerful additional constraint on any comparison with simulations.

We simulated \textit{Athena} X-IFU spectra for a 50~ks observation of NGC~6338, using the available response matrices and simulated background files\footnote{available at http://x-ifu.irap.omp.eu/resources-for-users-and-x-ifu-consortium-members/}, basing the input models on our \chandra\ measurements. For a region of area 1~arcmin$^2$ in the shock region west of VII~Zw~700, with mean kT=4.25~keV, we found that the 6.7~keV He-like Fe line was clearly detected, and redshift could be measured to an accuracy of $\sim$50\kmps (3$\sigma$ uncertainty). For 5\arcs-radius regions in the nucleus of NGC~6338 and the west knot of the north core, we found that even when using two-temperature models with different redshifts, we were able to recover the input values to accuracies of $<$30\kmps. The numerous emission lines from the 1-2~keV gas in these regions allows the mean redshift to be determined with great accuracy.

These simple tests suggest that \textit{Athena} will be able to measure the velocity field of key components of the NGC~6338 system. As well as providing constraints on any comparison with merger simulations, it will also allow a direct test of whether the gas of the north core is still associated with VII~Zw~700 or whether ram-pressure effects have pushed it back behind the galaxy along the line of sight. The spatial resolution of the X-IFU will be comparable to the scale of the hot gas filaments in the south core, so a comparison with the velocity of their denser H$\alpha$ counterparts will also be possible, as will a more detailed determination of whether the north tail is indeed affected by turbulent instabilities.

\section{Summary and Conclusions}
\label{sec:conc}

The NGC~6338 system provides an excellent example of both rapid radiative cooling and violent merger processes at the group scale. Prior studies have identified the two cool cores, their associated tails and cold fronts, the H$\alpha$ filaments and inner cavities in the south core, and noted that a significant component of the merger velocity is along the line of sight. In this paper we have presented results from a combination of deep \chandra\ and \xmms\ X-ray data, GMRT radio observations, and APO H$\alpha$ imaging and spectroscopy. We have used these to explore the dynamics of the system, the properties of the intra-group medium, and the highly structured cool cores with their associated dominant galaxies. We summarise our conclusions below:

\begin{enumerate}
\item We have used \chandra\ and \xmms\ spectral maps to show the temperature, abundance, pressure and entropy structure of the system. We confirm that the two cool cores are associated with cool ($\sim$2~keV), enriched (0.75\Zsol), low entropy tails extending to north and south. The gas between and to east and west of the cores appears to have been shock heated by the merger, reaching temperatures of $\sim$5~keV in some regions. This shocked material has low abundances (0.2-0.4\Zsol), suggesting it originated outside the enriched central regions of the progenitors. A pseudo-pressure map shows high pressures in the cores and the shocked region between them, with the high pressure region having the greatest east-west extent in the shocked region. We find that the surface brightness and temperature distributions are in good agreement with predictions from simulated mergers of similar-mass systems merging along the line of sight. We confirm the presence of cold fronts on the facing edges of the two cores, but do not find evidence of shock fronts; this is expected, since the shock direction will be close to the line of sight. Based on a comparison of the peak temperature with the range of temperature expected in the pre-shock gas, we find that likely Mach numbers for the shocks of $\mathcal{M}$=2.3-3.1, and likely merger velocities of 1700-1800\kmps. This suggests that the merger is amongst the most violent yet observed between galaxy groups.

\item We use a Gaussian mixture model to assess the 3-dimensional (RA, Dec, velocity) distribution of galaxies within 1.5~Mpc and 2000\kmps\ of the group center. We find that the galaxy population divides into two subsets, one centred close to the velocity of NGC~6338 with a narrow velocity dispersion ($\sigma$=261$\pm$31\kmps), the other including VII~Zw~700 and having a broader dispersion ($\sigma$=496$\pm$48\kmps). The two subsets are relatively evenly distributed in the plane of the sky, but galaxies with the highest recession velocities tend to be located to the north and west, consistent with their association with VII~Zw~700 and the northern cool core. The mean velocity difference between the two subsets is $\sim$850\kmps, but the difference between the two dominant galaxies is $\sim$1400\kmps. The latter is consistent with the merger velocity estimated from the X-ray temperatures. 

\item Our H$\alpha$ imaging confirms the previously detected 3-branched filamentary nebular emission in NGC~6338, and traces some diffuse extension to the northeast and southwest. We find a total H$\alpha$ luminosity of 3.1$\pm$0.6$\times$10$^{40}$\ergps, consistent with previous measurements. We also identify a previously unknown H$\alpha$ blob in the northern core, with a luminosity of 3.1$\times$10$^{39}$\ergps\ and a recession velocity consistent with that of VII~Zw~700. In both cores, the H$\alpha$ emission is associated with dense, cool, X-ray structures. In the south core these follow the same branching filamentary structure as the H$\alpha$ gas, with temperatures $\sim$1~keV and low entropies (5-8\kevcmsq) and cooling times (130-200~Myr). The extent of the filaments matches the radius at which the isochoric cooling time falls below 1~Gyr, and is similar to the radius at which t$_{cool}$/t$_{ff}$ reaches its minimum value (9.7$\pm$3.2). The central cooling time in the core is only 63$\pm$7~Myr. In the north core, the coolest gas forms a bar running roughly southeast-northwest, with the H$\alpha$ associated with a dense knot at the west end. However, cooling times are $<$1~Gyr throughout the core, and the gas appears offset from the dominant galaxy. This suggests that the motion of the core through the surrounding medium has pushed the gas back, perhaps disrupting the connection between cooling and the AGN. Comparing the two cores demonstrates the differing degree of impact a merger may have on cool cores; whereas the north core has been significantly disturbed by its infall, the larger south core has as yet been only minimally affected by the interaction.

\item We identify the same features previously identified as cavities near the centre of NGC~6338, and use our deeper \chandra\ data to more accurately estimate their shape. We also identify a possible cavity at larger radius in the south core, and a possible pair of cavities in the north core. We estimate the enthalpy and dynamical timescales of the cavities, and find that the implied jet powers are capable of balancing cooling in the two cores, although only marginally in the north core. In the south core, while the inner cavities lie to either side of the cooling filaments, the outer cavity lies at their end, implying that if all cavities are real AGN-inflated bubbles, the AGN jet axis must have changed over the course of $\sim$15~Myr. The presence of the outer cavity at the end of the filaments suggests that its rise could have been responsible for their formation, through the uplift of cool gas from the core. The cavity is probably too small to have uplifted the whole filament structure, but might have lifted enough gas to trigger further cooling.

\item Under the assumption of hydrostatic equilibrium, we estimate the total mass of the system to be $M_{tot,200}$=1.10$^{+0.02}_{-0.01}$$\times$10$^{14}$\Msol, in reasonable agreement with the Planck SZ mass estimate, and the gas mass to be 1.82$\pm$0.01$\times$10$^{13}$\Msol, within a radius R$_{200}$=980~kpc. However, while the statistical uncertainties on these estimates are small, the choice of whether to include the outermost bin of our deprojected temperature profile in the analysis alters the mass by a factor $\sim$1.6. Given the disturbed state of the system and evidence of relatively violent shock heating, more detailed analysis of the mass profile is probably impractical. 

\item We measure the radio flux density of the AGN of NGC~6338 at three frequencies (147, 333 and 1388~MHz) using archival GMRT data and, in combination with literature results, measure its spectral index to be $\alpha^{8400}_{74}$=0.38$\pm$0.03. We see a hint of extension to the southwest in the deepest observation, at 1.39~GHz, but do not observe the cross-shaped extension reported by Wang et al. from a 1.4~GHz VLA snapshot. We find the AGN to be X-ray faint, and place a 3$\sigma$ upper limit on its luminosity of L$_{2-10~keV}$$\leq$1.35$\times$10$^{40}$\ergps. We do not identify any radio sources associated with the dominant galaxy pair of the northern core, VII~Zw~700, but do find evidence of an X-ray AGN in the larger galaxy, MCG+10-24-117, with luminosity L$_{2-10~keV}$=4.17$^{+1.64}_{-1.68}$$\times$10$^{39}$\ergps. We place an upper limit on the AGN luminosity of the smaller galaxy. 

\end{enumerate}

\medskip
\noindent\textbf{Acknowledgments}\\
We thank the anonymous referee for their comments on the paper.
E.~O'S. thanks A.~J.~R. Sanderson for the original suggestion that NGC~6338
deserved investigation, and gratefully acknowledges the support for this
work provided by the National Aeronautics and Space Administration (NASA)
through \chandra\ Award Number GO7-18162X issued by the \chandra\ X-ray
Center, which is operated by the Smithsonian Astrophysical Observatory for
and on behalf of the National Aeronautics Space Administration under
contract NAS8-03060. M.~S. acknowledges support from the Chandra Award
AR7-18016X and NSF grant 1714764. This research is based in part on
observations obtained with \xmm, an ESA science mission with instruments
and contributions directly funded by ESA member states and NASA, and with
the GMRT, which is a national facility operated by the National Center for
Radio Astrophysics (NCRA) of the Tata Institute for Fundamental Research
(TIFR). This research has made use of data based on observations obtained
with the Apache Point Observatory 3.5-meter telescope, which is owned and
operated by the Astrophysical Research Consortium. APO observations were
supported by the F. H. Levinson Fund of the Silicon Valley Community
Foundation through a donation to the University of Virginia. This
research has made use of the NASA/IPAC Extragalactic Database (NED) which
is operated by the Jet Propulsion Laboratory, California Institute of
Technology, under contract with NASA. We acknowledge the usage of the
HyperLeda database (http://leda.univ-lyon1.fr). This work made use of data
from the Galaxy Cluster Merger Catalog (http://gcmc.hub.yt).


\bibliographystyle{mnras}
\bibliography{../paper}

\label{lastpage}

\end{document}